\documentclass[twoside,12pt]{article}
\pdfoutput=1
\usepackage{graphicx}
\usepackage{epsfig}
\usepackage[utf8]{inputenc}
\usepackage{url}
\usepackage[cal=esstix]{mathalfa}
\usepackage{amsmath}
\usepackage{amsthm}
\usepackage{amssymb} 
\usepackage{float, placeins}
\usepackage{caption, subcaption}
\usepackage{dblfloatfix,fixltx2e}
\usepackage[font={it}]{caption}
\usepackage[margin=2.0cm]{geometry}
\usepackage{cancel}
\usepackage{xfrac}
\usepackage{xspace}
\usepackage{hyperref}
\usepackage{booktabs}
\usepackage[binary-units=true]{siunitx}
\usepackage{accents}
\usepackage{titling}
\usepackage[super]{nth}
\usepackage[toc,page]{appendix}
\usepackage{algorithm}
\usepackage{algpseudocode}
\usepackage{bm} % bold math symbols
\usepackage{authblk}
\newcommand{\tev}{\tera\electronvolt}
\newcommand{\gev}{\giga\electronvolt}
\newcommand{\mev}{\mega\electronvolt}
\newcommand{\mm}{\milli\meter}
\newcommand{\kNN}{\ensuremath{k\text{NN}}\xspace}

\begin{document}

\title{Muon Energy Measurement from Radiative Losses in a Calorimeter for a Collider Detector}

\author[1]{Tommaso Dorigo}
\author[2]{Jan Kieseler}
\author[3]{Lukas Layer}
\author[4]{Giles C. Strong}

\affil[1,3,4]{INFN, Sezione di Padova}
\affil[2]{CERN}
\affil[3]{Università di Napoli ``Federico II"}
\affil[4]{University of Padova}

\date{August 25, 2020}

\maketitle
\begin{abstract}
The performance demands of future particle-physics experiments investigating the high-energy frontier pose a number of new challenges, forcing us to find new solutions for the detection, identification, and measurement of final-state particles in subnuclear collisions. One such challenge is the precise measurement of muon momenta at very high energy, where the curvature provided by conceivable magnetic fields in realistic detectors proves insufficient to achieve the desired resolution.

In this work we show the feasibility of an entirely new avenue for the measurement of the energy of muons based on their radiative losses in a dense, finely segmented calorimeter. This is made possible by the use of the spatial information of the clusters of deposited photon energy in the regression task. Using a homogeneous lead-tungstate calorimeter as a benchmark, we show how energy losses may provide significant complementary information for the estimate of muon energies above 1 TeV.
\end{abstract}
%\markright{Customized 

\clearpage
\section{Introduction \label{s:introduction}}

Muons have been excellent probes of new phenomena in particle physics ever since their discovery in cosmic showers\cite{muon_disc1, muon_disc2}. Their detection and measurement enabled groundbreaking discoveries, from those of heavy quarks\cite{richter,lederman,topdisc} and weak bosons\cite{rubbia} to the discovery of the Higgs boson\cite{higgsatlas,higgscms}; most recently, a first evidence for $H \to \mu \mu$ decays has been also reported by CMS\cite{hmmcms}, highlighting the importance of muons for searches as well as measurements of standard model parameters. This is due to their particular physical properties, most notably their interaction with matter. With a mass 200-times higher than that of the electron, muons lose little energy by electromagnetic radiation and behave as minimum ionizing
particles in a wide range of energies, making them easily distinguishable from long-lived light hadrons such as charged pions and kaons. 

The peculiar properties of muons will remain of high value for new physics searches in future high-energy colliders, where their identification is typically easier than that of electrons of the same energy. A number of heavy particles predicted by new-physics models are accessible preferentially, and in some cases exclusively, by the detection of their decay to final states that include electrons or muons; in particular, the reconstruction of the resonant shape of dileptonic decays of new gauge-bosons $Z'$ resulting from the addition of an extra U(1) group or higher symmetry structures to the Standard Model\cite{zprime1,zprime2} constitutes a compelling reason for seeking the best possible energy resolution for electrons and muons of high energy.
\begin{center}
\begin{figure}[h!]
\centerline{\includegraphics[width=0.9\textwidth]{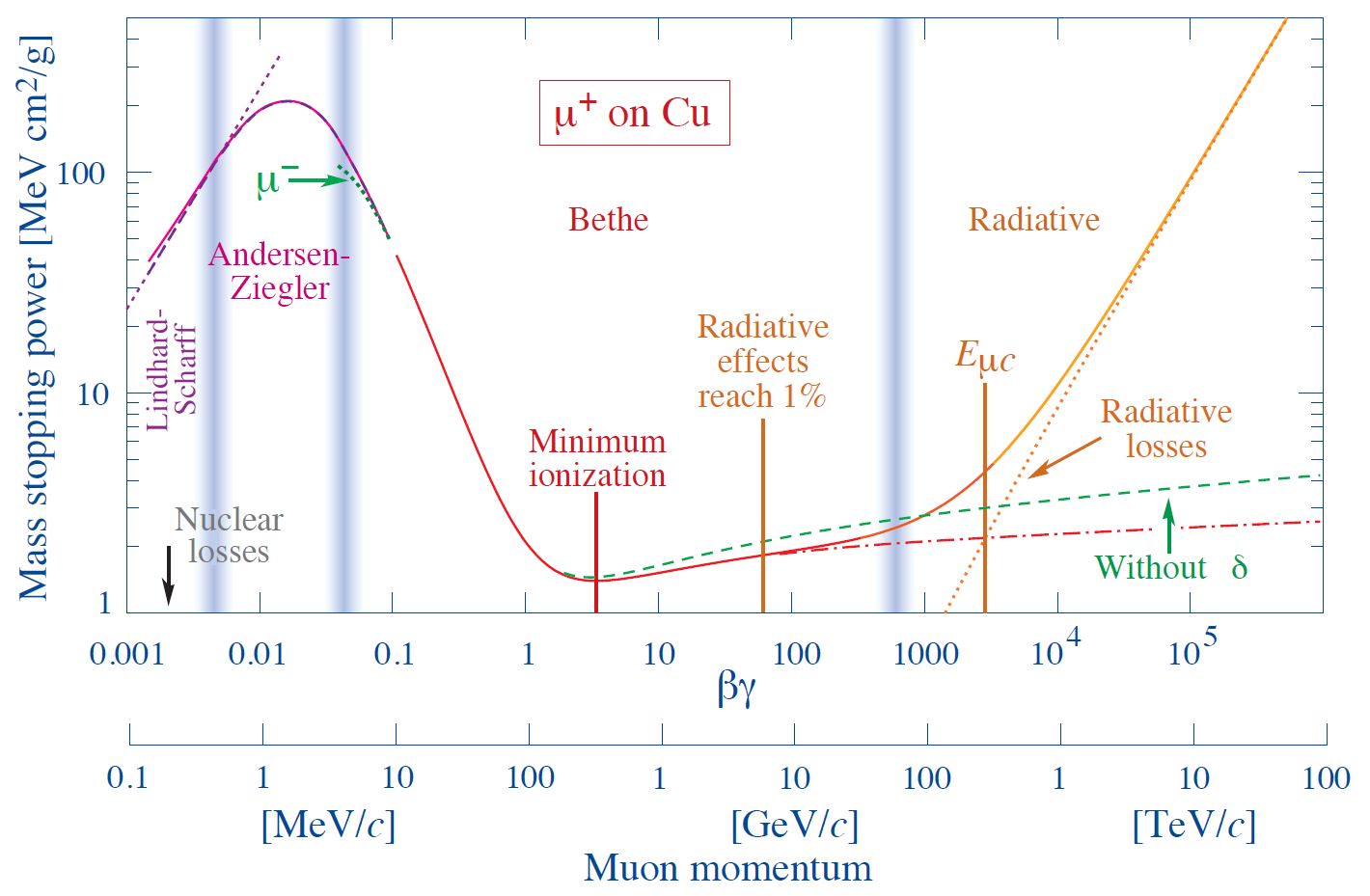}}
\caption{\em Mass stopping power for muons in the \SI{1}{\mev} to  \SI{1000}{\tev} range, in \si{\mev\centi\meter^2\per\gram}. The rise in radiative loss becomes important above \SI{100}{\gev} (reprinted from the Review of Particle Properties\cite{pdg}.)}
\label{f:muonEloss}
\end{figure}
\end{center}

\noindent
Unfortunately, the very features that make muons special and easily distinguishable from backgrounds also hinder the precise measurement of their energy. While the energy of electrons is effectively
inferred from the electromagnetic showers they initiate in dense calorimeters, muon energy estimates rely on a measurement of their momentum through precise tracking over high B field integrals. If we consider the ATLAS and CMS detectors as reference points, 
we observe that the relative resolution of muon transverse momentum achieved in those state-of-the-art instruments ranges from 8 to 20\% (ATLAS) or 6 to 17\% (CMS) at \SI{1}{\tev}\cite{atlasmuon,cmsmuon}, depending on detection and reconstruction details.
%; by comparison, for electrons the resolution ranges from XX to YY\% (ATLAS) and from XX to YY\% (CMS) at the same energy\cite{atlaselec,cmselec}. 
Clearly, calorimetric measurements win over curvature determinations at high energy for not minimum-ionizing particles, due to the different scaling properties of the respective resolution functions (linear in energy for curvature-driven ones, while depending on $1/\sqrt{E}$ for calorimetric ones).

In truth, muons do not behave as minimum-ionizing particles at high energy: rather, they show a relativistic rise of the energy loss\cite{pdg} above roughly \SI{100}{\gev} (see \autoref{f:muonEloss}). The effect is clear, but it is undeniably very small in absolute terms. For that reason, to our knowledge it has never been exploited to estimate the muon energy. That is what this work aims to show: an attractive avenue for the energy measurement of muons of very high energy is possible in a sufficiently thick and fine-grained calorimeter, if tailored machine-learning reconstruction methods are employed for the task. The input of such a measurement is the pattern of detected energy deposition of soft photons radiated from the energetic muon track.

In \autoref{s:detector} below we show the idealized calorimeter we have employed for the goal of this study. In \autoref{s:features} we proceed to describe the high-level features we extract from the energetic and spatial information of each muon interaction event; these features are used as an input for the regression task. In \autoref{s:regressor} we discuss the machine learning tool we used for the regression of muon energy from the measured energy deposits. In \autoref{s:tests} we detail our results. We offer some concluding remarks in \autoref{s:conclusion}.

\section{Detector geometry and simulation}\label{s:detector}

Since our goal in this work is to show the feasibility of muon-energy estimation from energy deposition in a calorimeter, we strip the problem of any complication from factors that are ancillary to the task. For that reason, we consider a homogeneous lead tungstate cuboid calorimeter,
with a total depth in $z$ of $\SI{2032}{\mm}$, corresponding to 10 $\lambda_0$, and a spatial extent of $\SI{120}{\mm}$ in $x$ and $y$. The calorimeter is segmented in 50 layers in $z$, each with a thickness of $\SI{39.6}{\mm}$, which corresponds to 4.5 radiation lengths, such that electromagnetic showers are well resolvable. A layer is further segmented in $x$ and $y$ in $32 \times 32$ cells, with a size of $\SI{3.73}{\mm} \times \SI{3.73}{\mm}$. This results in 51,200 channels in total.

\begin{figure}[h!]
\centering
\includegraphics[width=0.8\textwidth]{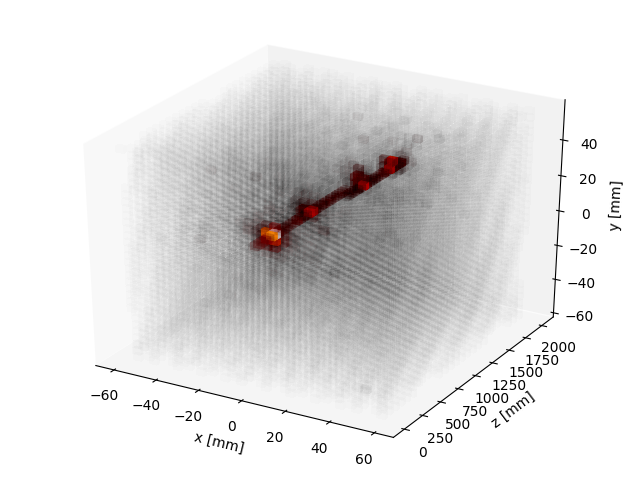}
\caption {\em Muon entering the calorimeter in $z$ direction. The colour palette indicates logarithmic energy deposits of a muon with an energy of approximately $\SI{290}{\gev}$. Black corresponds to zero, red to intermediate, and white to the maximum energy.}
\label{f:geometry}
\end{figure}

\subsection {The simulated data}

We generate muons at $z=-\SI{50}{\mm}$ with respect to the calorimeter front face with a momentum in $z$ direction between $\SI{100}{\gev}$ and $\SI{2}{\tev}$ as it is the relevant energy range for this study. The muon position in $x$ and $y$ is randomly chosen within $|x|\leq\SI{20}{\mm}$ and $|y|\leq\SI{20}{\mm}$. The momentum components in $x$ and $y$ direction are set to zero. The interaction of the muons with the detector material is simulated using Geant4~\cite{GEANT4}.  In total 411,000 events are generated, out of which 396,000 are used for training and 15,000 for testing; of the latter, 3000 are used to extract parameters of a bias correction function, and 12,000 are studied to gauge the regression performance.

The detector geometry, as well as a possible radiation pattern of a muon entering the calorimeter is shown in \autoref{f:geometry}. Even at a relatively low energy of $\SI{290}{\gev}$, the radiation pattern is clearly visible.

%%%%%%%%%%%%%%%%%%%%%
\clearpage
\section{Event features \label{s:features}}

The regression task we set up in Sec.\autoref{s:regressor} uses 16 event features extracted from the spatial and energy information of calorimeter cells. In this section we describe those inputs.

Some of the features describe general properties of the energy deposition ({\em e.g.}, the sum of all deposited energy above or below a $E=\SI{0.1}{\gev}$ threshold), some are partly reliant on fine-grained information (moments of the energy distribution, in five regions of detector depth: $z<\SI{400}{\mm}$, $400<z<\SI{800}{\mm}$, $800<z<\SI{1200}{\mm}$, $1200<z<\SI{1600}{\mm}$, and $z>\SI{1600}{\mm}$; and imbalance of the deposited energy in the transverse plane), and a few more are instead derived from a full-resolution clustering of the energy deposition pattern, which is briefly described below.

\subsection{The clustering algorithm \label{s:cluster}}

We set a minimum threshold of $E_{\text{thr.}}=\SI{0.1}{\gev}$ for the seed tower of the clustering. The search for clusters starts with calorimeter towers in the column of towers (of same transverse coordinates $x$ and $y$) traversed by the incoming muon\footnote{Although the identity of the correct hit point is unequivocal in our setup, since muons are simulated to impinge normally on the detector face and the corresponding column of calorimeter towers is always the one with the largest energy deposition, this assumption is made under the hypothesis that a previous device has identified the incoming track position at the calorimeter face.}, and performs the following calculations:\par

\begin{enumerate}
    \item The highest-energy tower is selected as a seed if it has $E>E_{\text{thr.}}$ and it lies along the muon direction;
    \item The six calorimeter towers adjacent in either $x$, $y$, or $z$ to the seed tower are added to the cluster if they have recorded a non-null energy deposition;
    \item Towers with non-null energy deposition are progressively added to the cluster if they are adjacent to already included towers;
    \item The final cluster is formed when there are no more towers passing the above criteria; at that point, features such as the number of included towers and the total cluster energy are computed. 
    \item All towers belonging to the cluster are removed from the list of still unassigned towers;
    \item The algorithm returns to step 1) to form other clusters.
\end{enumerate}

\noindent
Once clusters seeded by the column of towers along the muon track are formed by the above procedure, a second set of clusters is sought for by looking for towers above threshold that remained unassigned to any cluster:\par

\begin{enumerate}
    \item The highest-energy tower still unassigned and above $E_{\text{thr.}}$ is considered, irrespective of its $x$, $y$ coordinates;
    \item Yet unassigned towers are added to the cluster if they are adjacent to it, according to the same prescription described above;
    \item Towers with non-null energy deposition are progressively added to the cluster if they are adjacent to towers already included;
    \item The final cluster is formed when there are no more towers passing the above criteria; features are then computed for the identified cluster;
    \item All towers belonging to the cluster are removed from the list of unassigned towers;
    \item The algorithm returns to step 1) to search for additional clusters.
\end{enumerate}

\noindent
Using the results of the above two-step clustering procedure, we form six summary statistics:\par

\begin{itemize}
    \item V[10]: The number of muon trajectory-seeded clusters;
    \item V[11]: The maximum total energy among those clusters;
    \item V[12]: The maximum number of cells among those clusters;
    \item V[13]: The number of clusters not seeded by a tower along the muon track;
    \item V[14]: The maximum number of cells among those clusters;
    \item V[15]: The maximum total energy among those clusters;
\end{itemize}

\subsection{Other features}\label{s:other_features}

We list below the other features we compute for each event:\par

\begin{itemize}
    \item V[0]: The total energy recorded in the calorimeter in towers above the $E_{\text{thr.}}>\SI{0.1}{\gev}$ threshold;
    \item V[1]: The total energy recorded in the calorimeter in all towers with energy below $E_{\text{thr.}}$;
    \item V[2]: the missing transverse energy deposition in the $xy$ plane, computed as $V[2]= \sqrt{E_x^2 + E_y^2}$, where $E_x=\sum{E_i \Delta x_i}$, $E_y=\sum{E_i \Delta y_i}$, and $\Delta x_i$, $\Delta y_i$ are the spatial distances in the transverse plane to the center of the tower hit by the muon; in this calculation, all towers are used;
    \item V[3]: This variable results from the same calculation extracting $V[2]$, but it is performed using only towers exceeding the $E_{\text{thr.}}=\SI{0.1}{\gev}$ threshold;
    \item V[4]: The second moment of the energy distribution around the muon direction in the transverse plane, computed with all towers as \par
    $V[4]= \sum_i [E_i (\Delta x_i^2 + \Delta y_i^2)]/ \sum_i E_i$,\\ 
    where indices run on all towers and the distances are computed in the transverse plane, as above;
    \item V[5]: The same as V[4], but computed only using towers located in the first \SI{400}{\mm}-thick longitudinal section of the detector along $z$;
    \item V[6]: The same as V[4], but computed only using towers in the $400<z_i<\SI{800}{\mm}$ region;
    \item V[7]: The same as V[4], but computed only using towers in the $800<z_i<\SI{1200}{\mm}$ region;
    \item V[8]: The same as V[4], but computed only using towers in the     $1200<z_i<\SI{1600}{\mm}$ region;    
    \item V[9]: The same as V[4], but computed only using towers in the $z_i\geq\SI{1600}{\mm}$ region.
\end{itemize}

\begin{figure}[h!]
    \begin{center}
        \begin{subfigure}[t]{0.45\textwidth}
            \begin{center}
                \includegraphics[width=\textwidth]{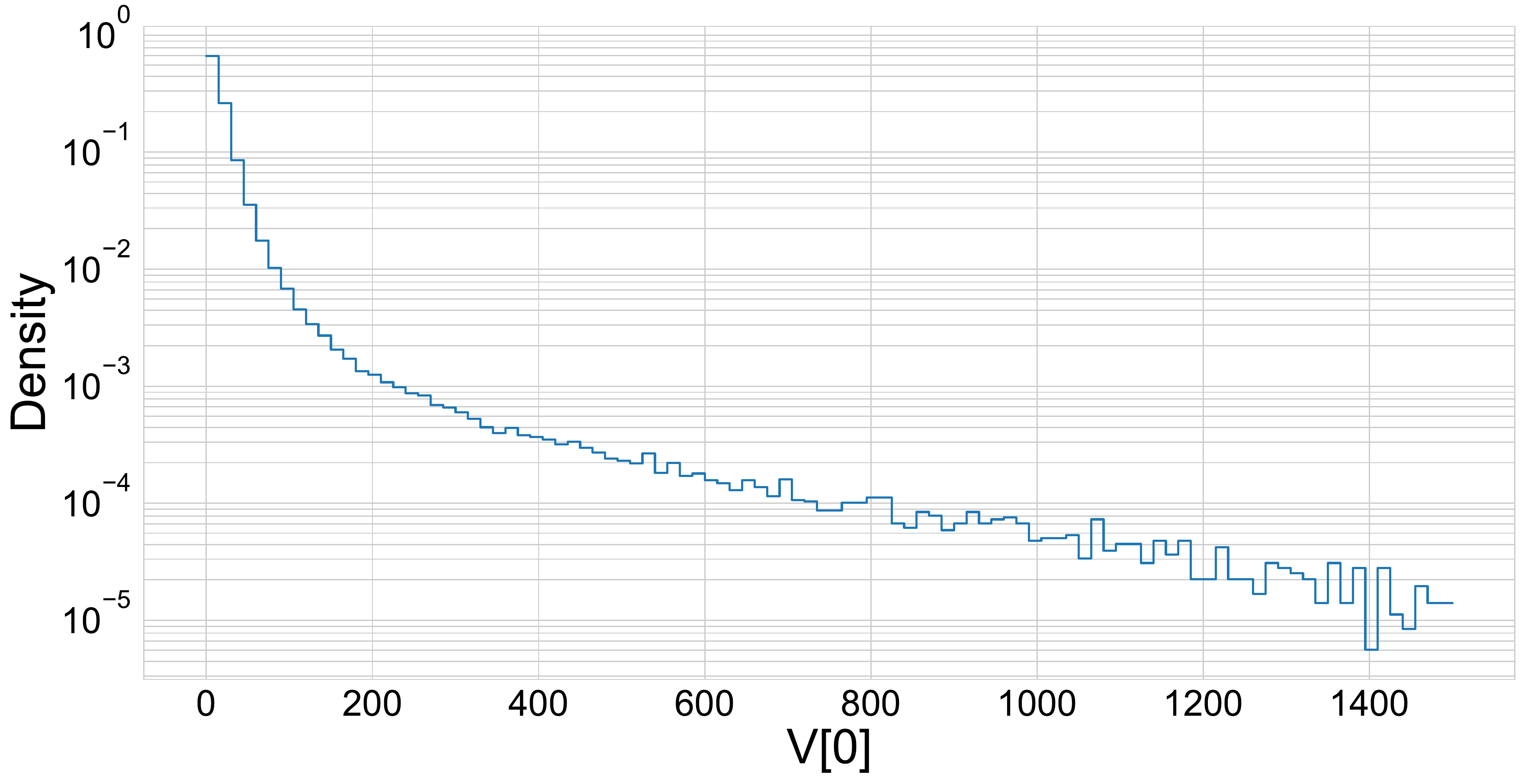}
            \end{center}
        \end{subfigure}
        \begin{subfigure}[t]{0.45\textwidth}
            \begin{center}
                \includegraphics[width=\textwidth]{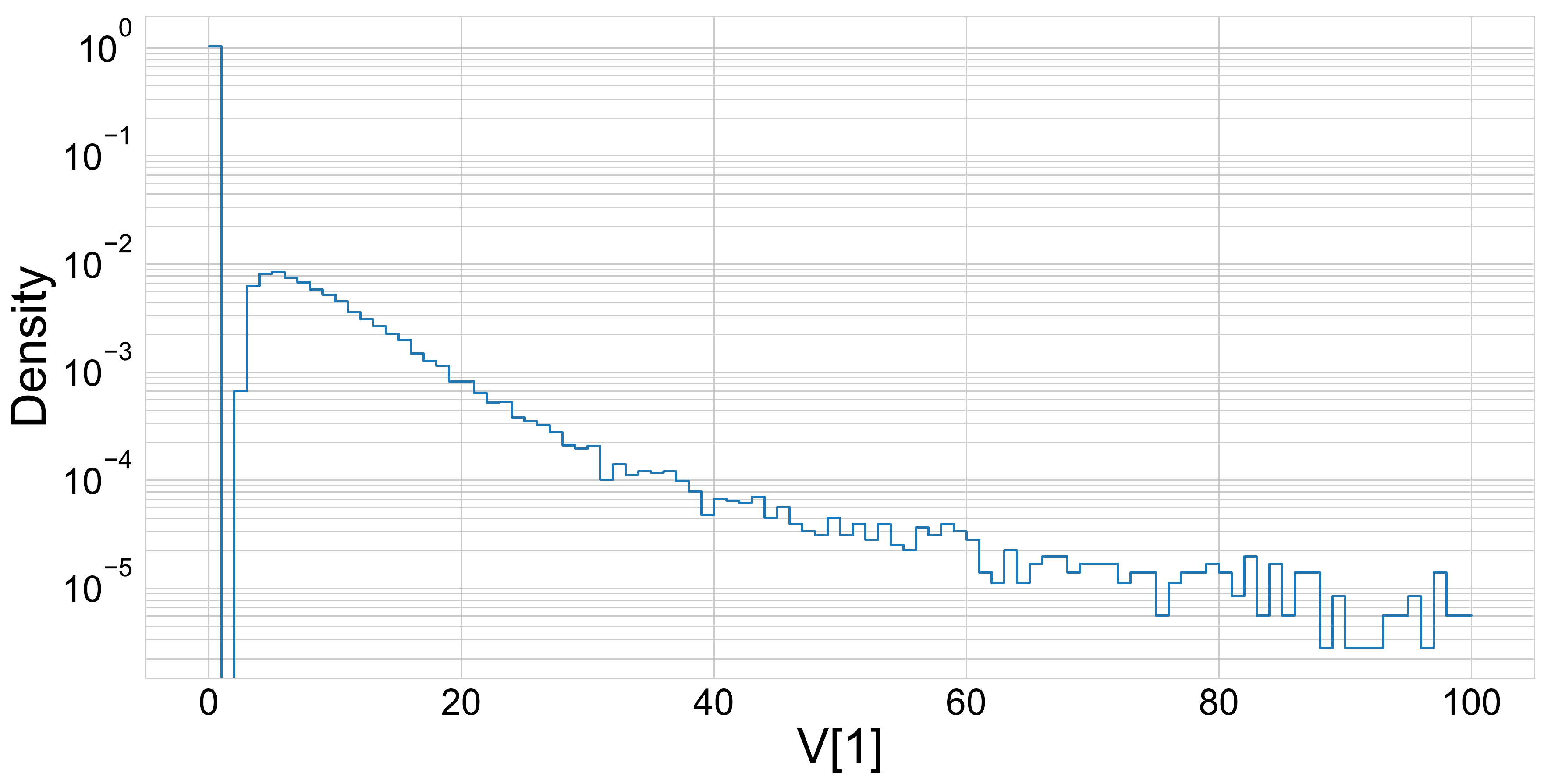}
            \end{center}
        \end{subfigure}
        \begin{subfigure}[t]{0.45\textwidth}
            \begin{center}
                \includegraphics[width=\textwidth]{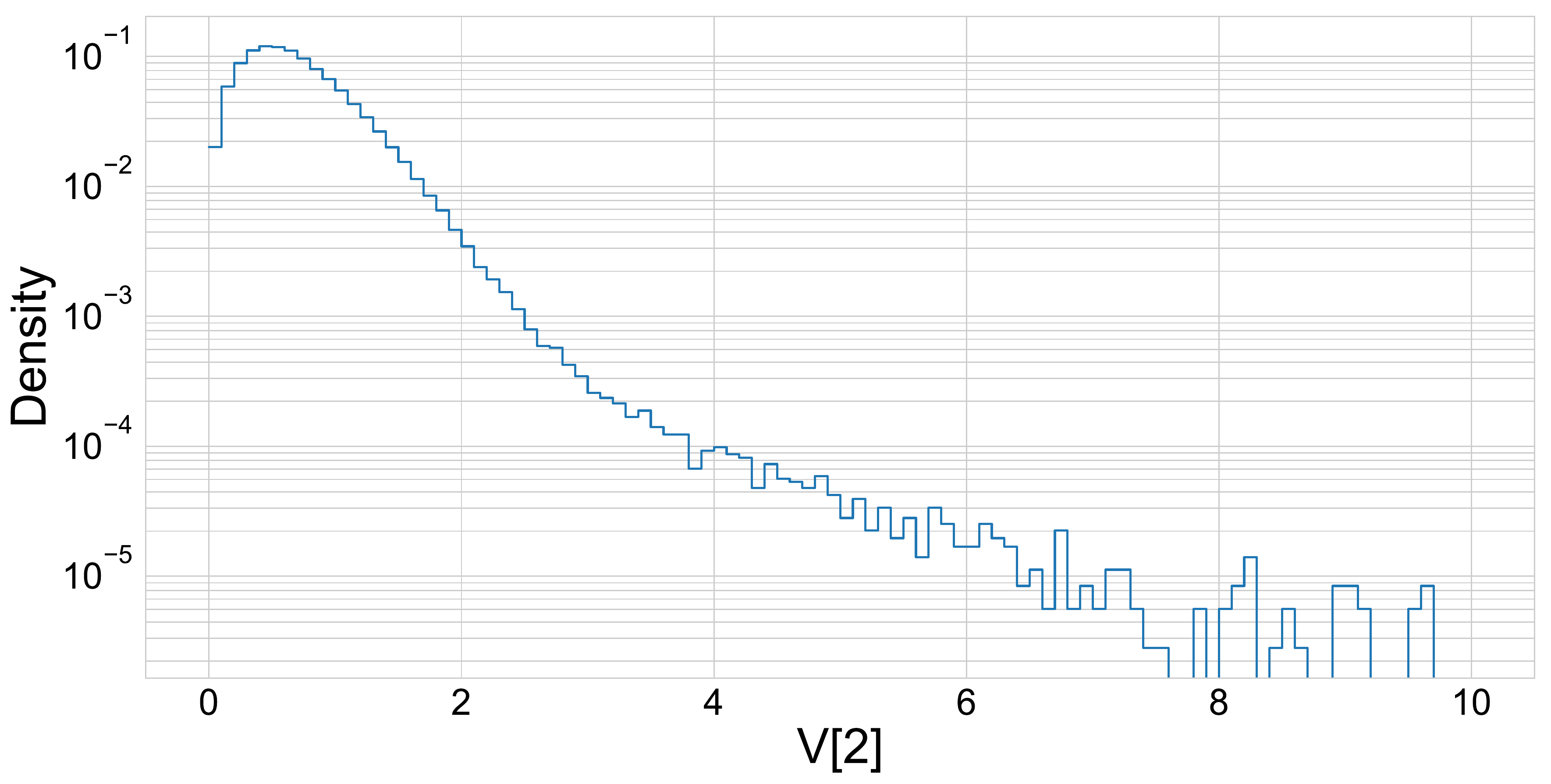}
            \end{center}
        \end{subfigure}
        \begin{subfigure}[t]{0.45\textwidth}
            \begin{center}
                \includegraphics[width=\textwidth]{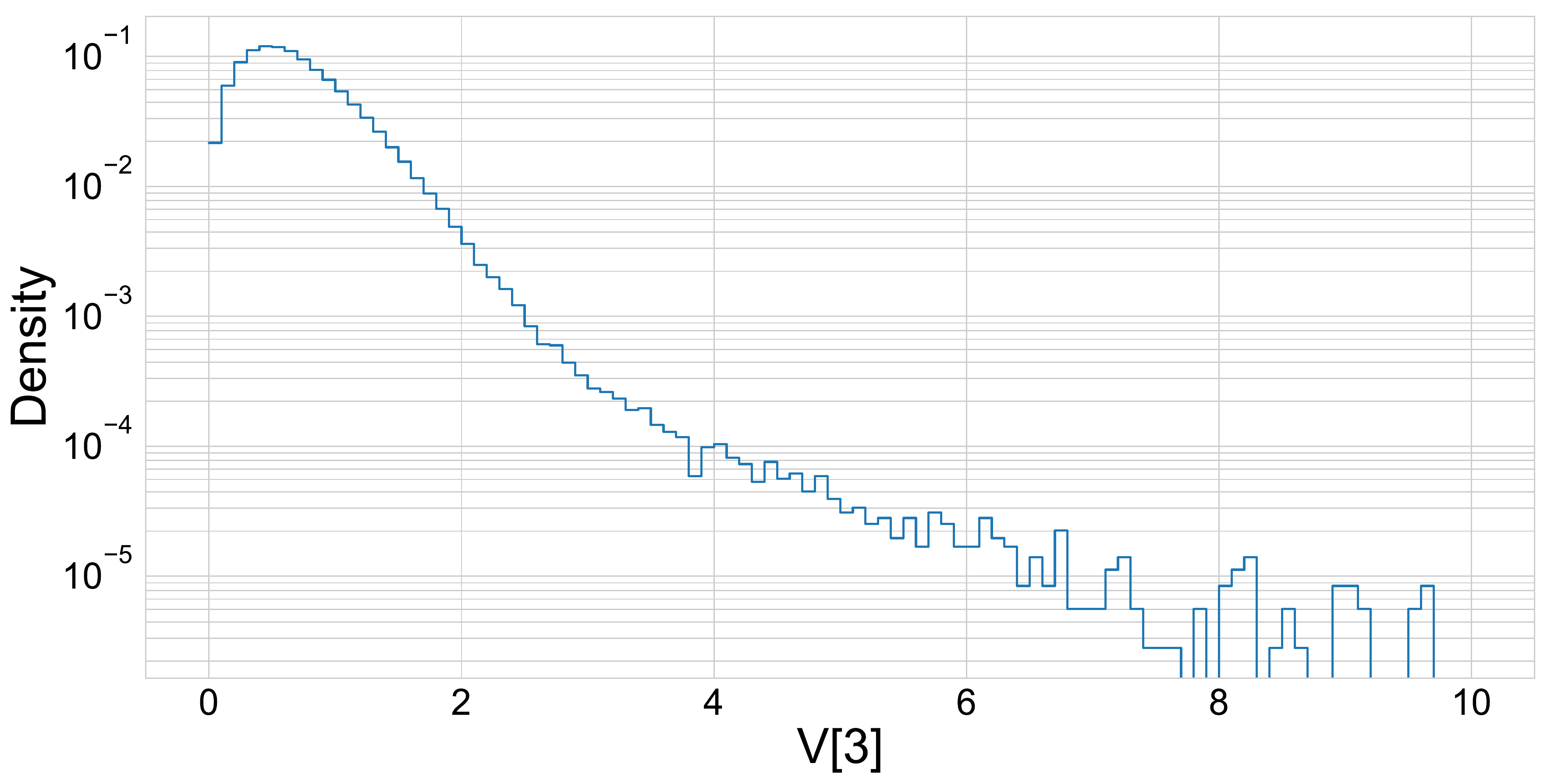}
            \end{center}
        \end{subfigure}
        \begin{subfigure}[t]{0.45\textwidth}
            \begin{center}
                \includegraphics[width=\textwidth]{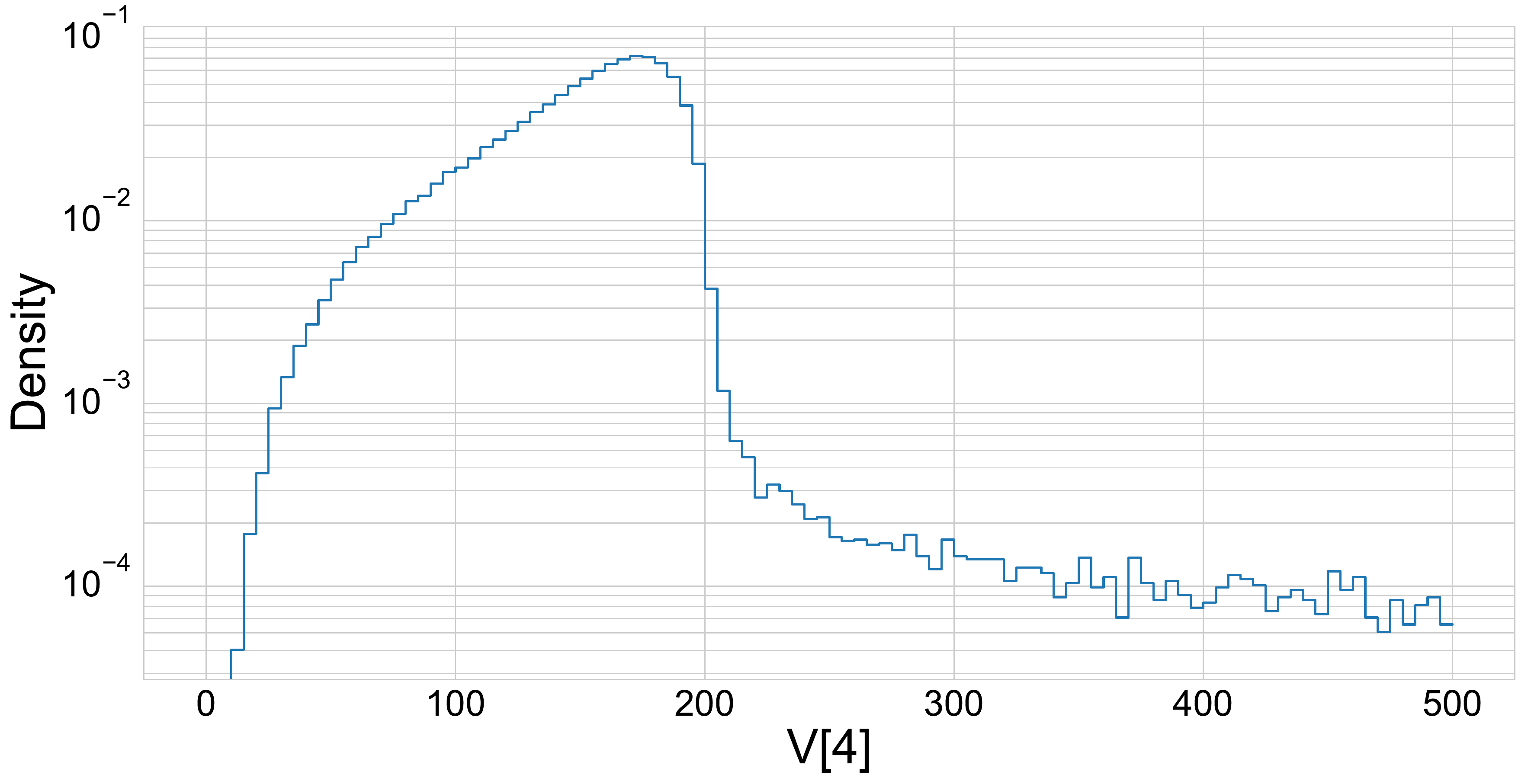}
            \end{center}
        \end{subfigure}
        \begin{subfigure}[t]{0.45\textwidth}
            \begin{center}
                \includegraphics[width=\textwidth]{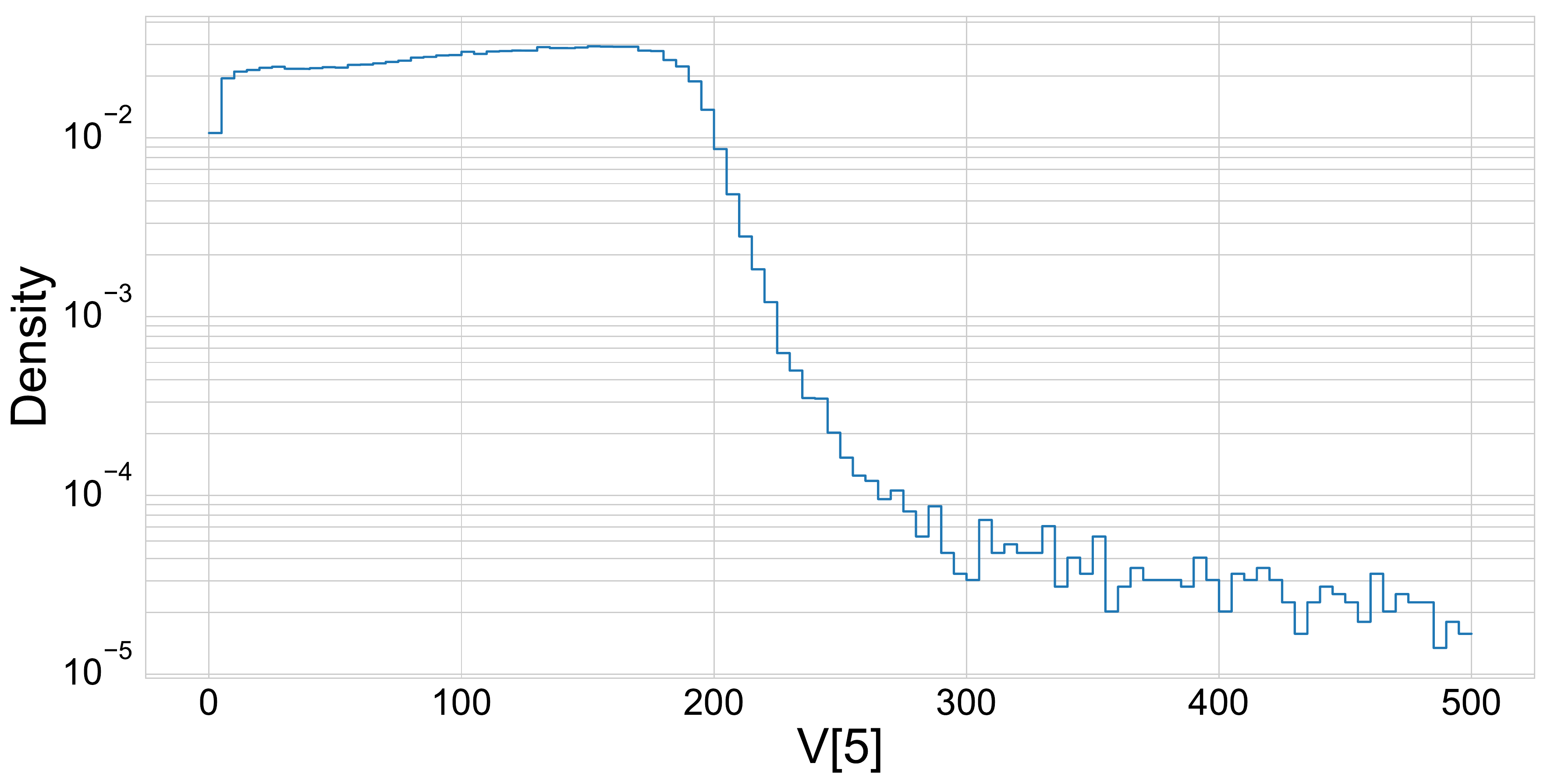}
            \end{center}
        \end{subfigure}
        \begin{subfigure}[t]{0.45\textwidth}
            \begin{center}
                \includegraphics[width=\textwidth]{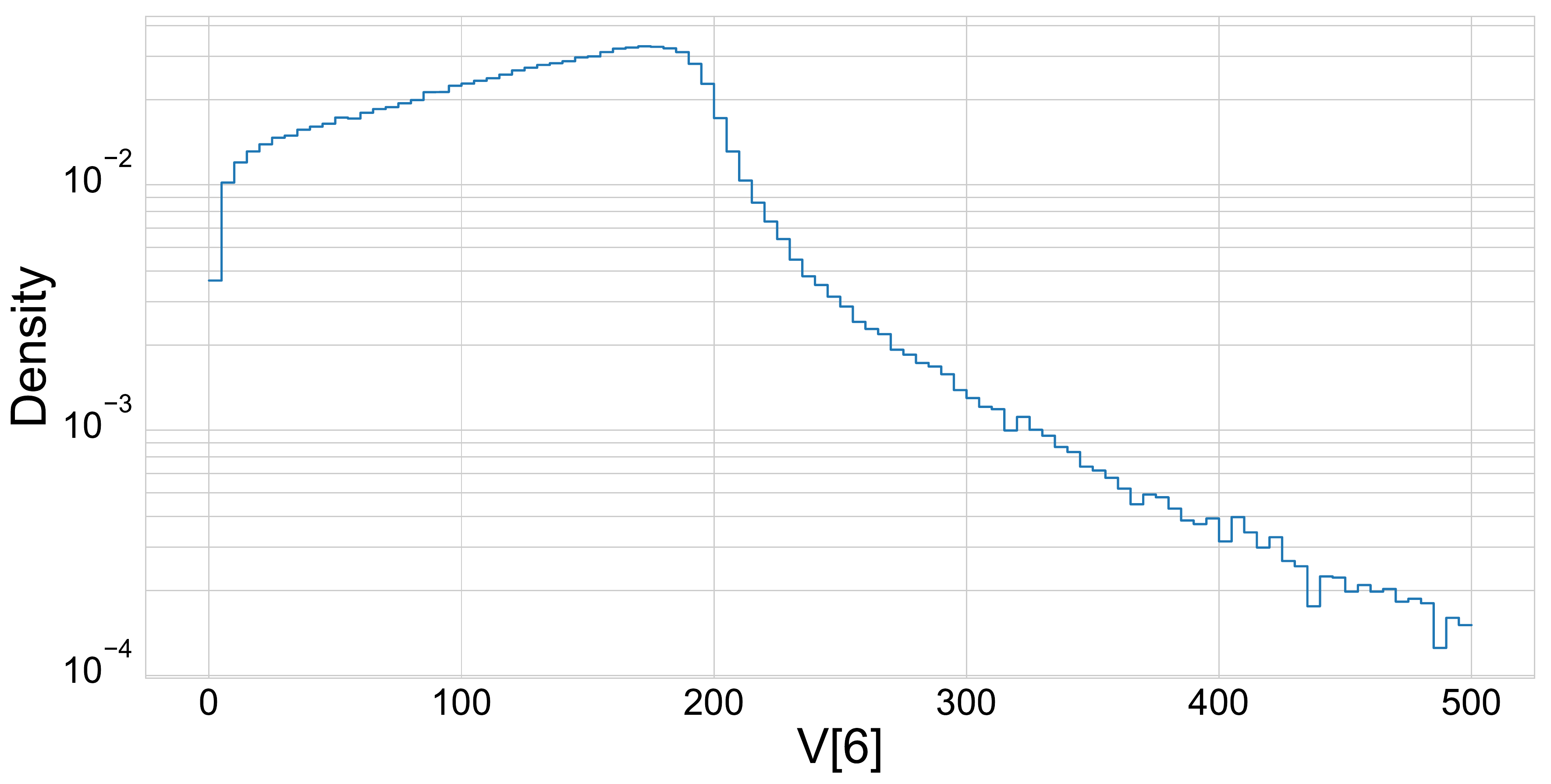}
            \end{center}
        \end{subfigure}
        \begin{subfigure}[t]{0.45\textwidth}
            \begin{center}
                \includegraphics[width=\textwidth]{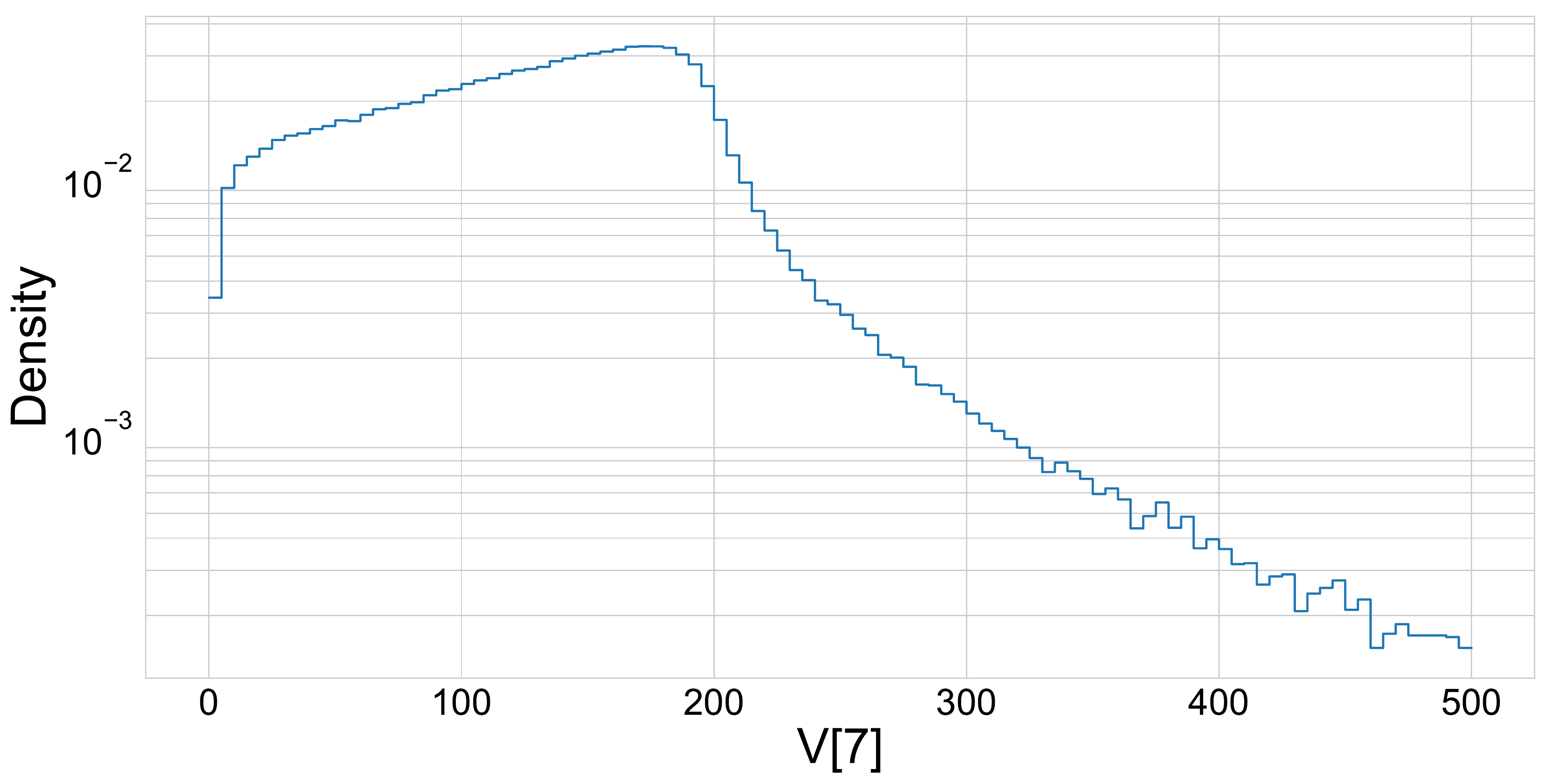}
            \end{center}
        \end{subfigure}
        \begin{subfigure}[t]{0.45\textwidth}
            \begin{center}
                \includegraphics[width=\textwidth]{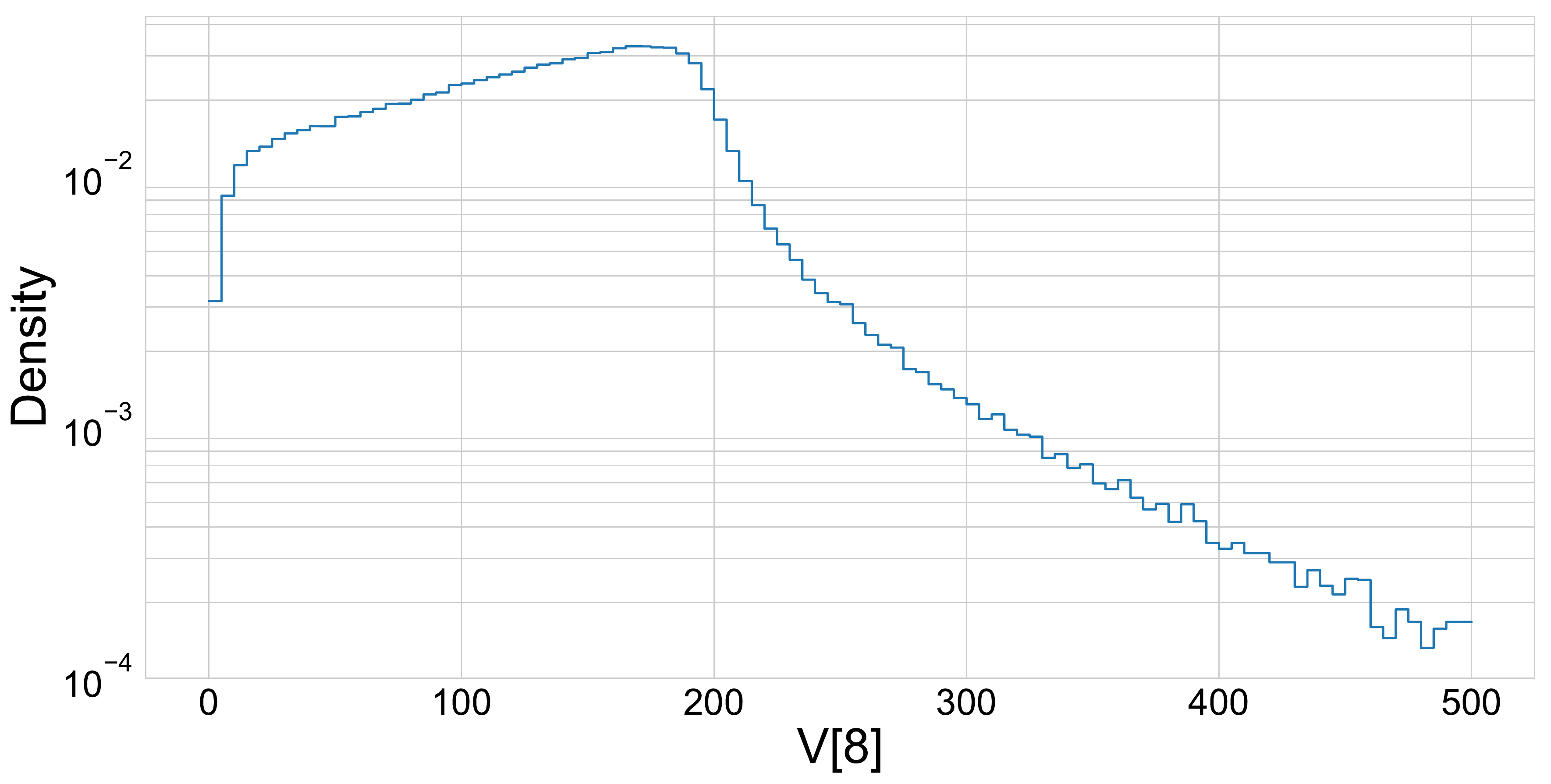}
            \end{center}
        \end{subfigure}
        \begin{subfigure}[t]{0.45\textwidth}
            \begin{center}
                \includegraphics[width=\textwidth]{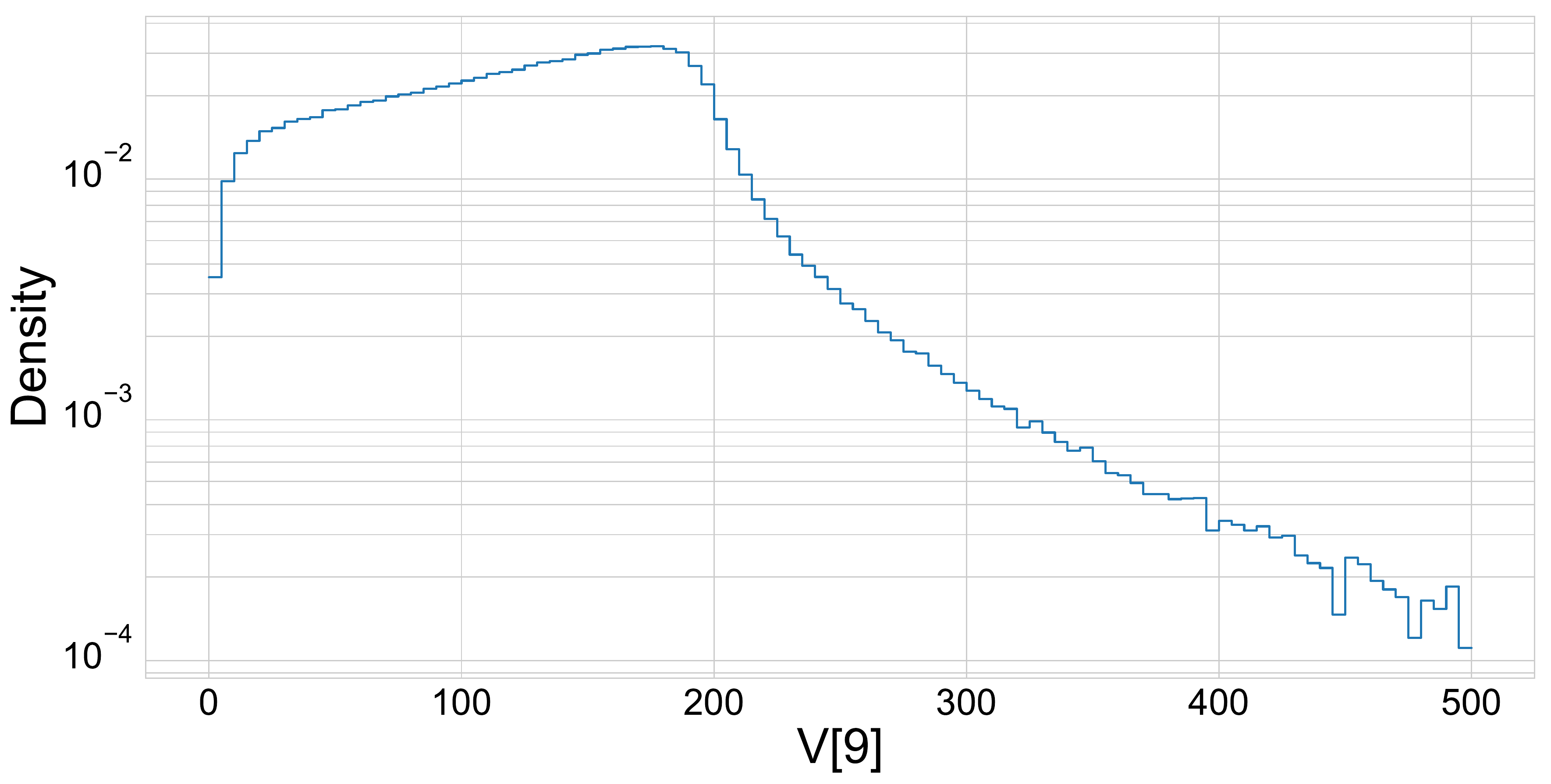}
            \end{center}
        \end{subfigure}
        \caption{\em Marginals of 10 non-cluster-related event features. Top row: V[0],V[1]; second row: V[2],V[3]; third row: V[4],V[5]; fourth row: V[6],V[7]; bottom row: V[8],V[9]. Features are defined in Section~\ref{s:other_features}.}
        \label{f:other_feats}
    \end{center}
\end{figure}

\begin{figure}[ht]
    \begin{center}
        \begin{subfigure}[t]{0.45\textwidth}
            \begin{center}
                \includegraphics[width=\textwidth]{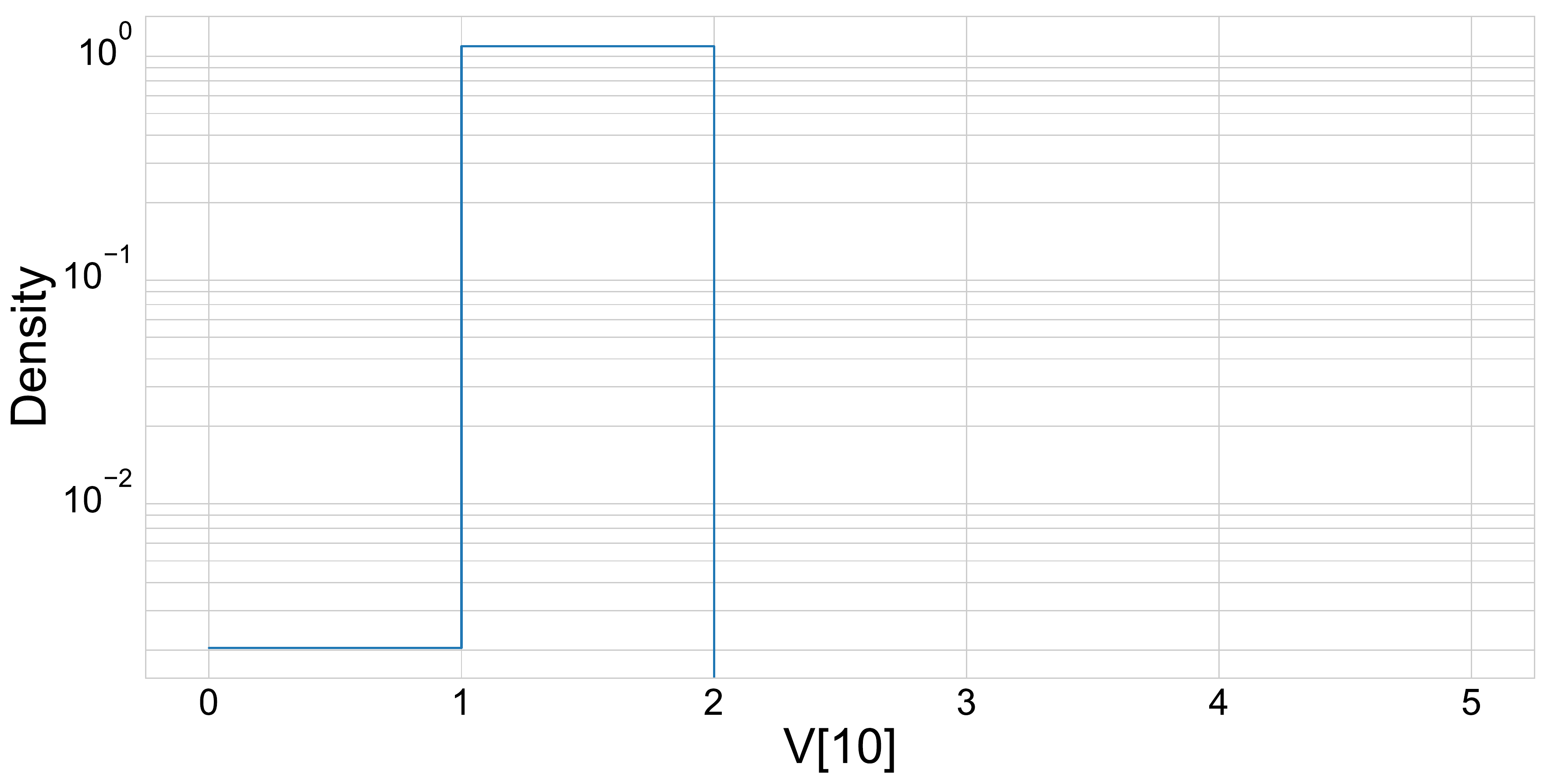}
            \end{center}
        \end{subfigure}
        \begin{subfigure}[t]{0.45\textwidth}
            \begin{center}
                \includegraphics[width=\textwidth]{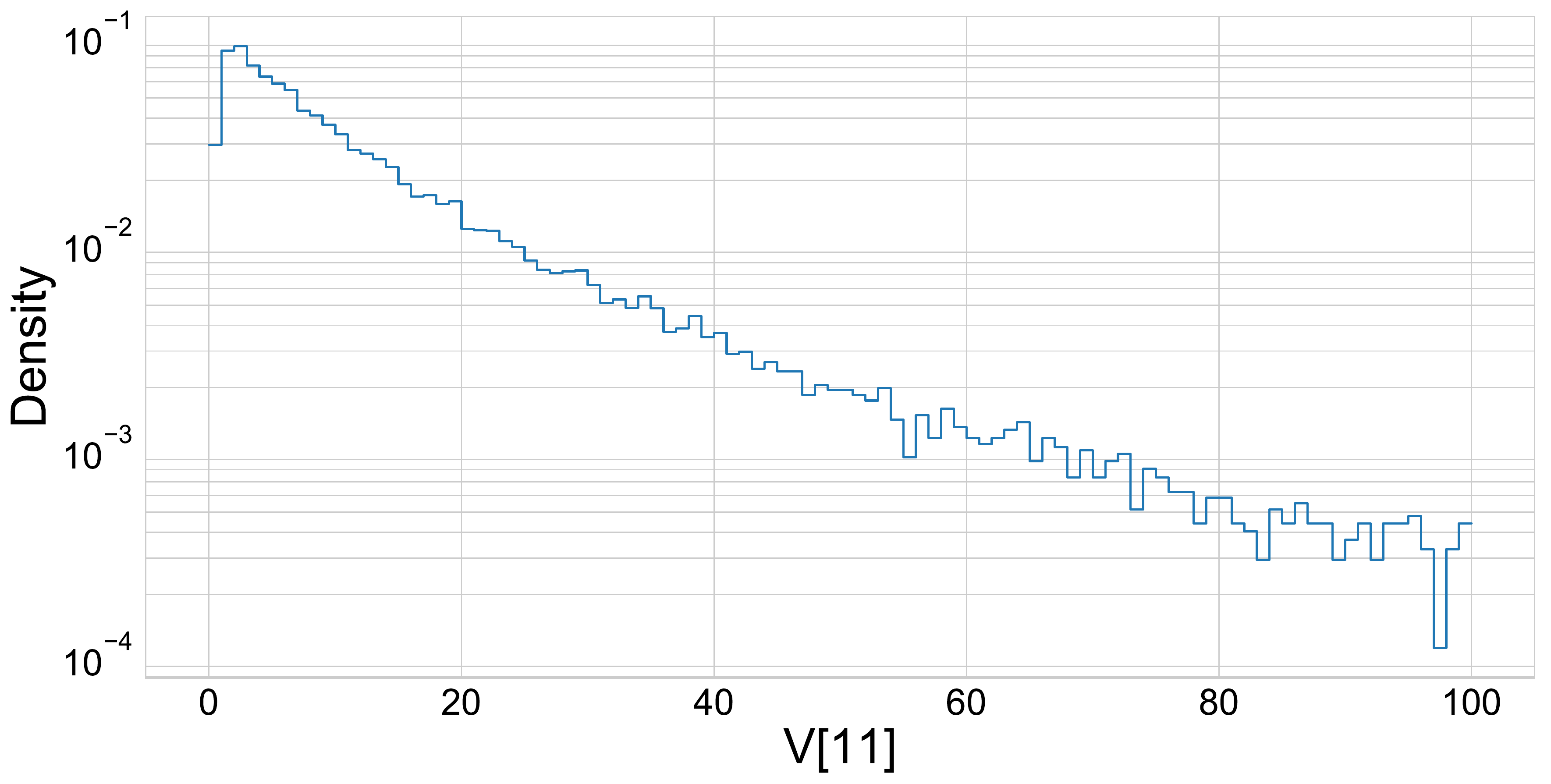}
            \end{center}
        \end{subfigure}
        \begin{subfigure}[t]{0.45\textwidth}
            \begin{center}
                \includegraphics[width=\textwidth]{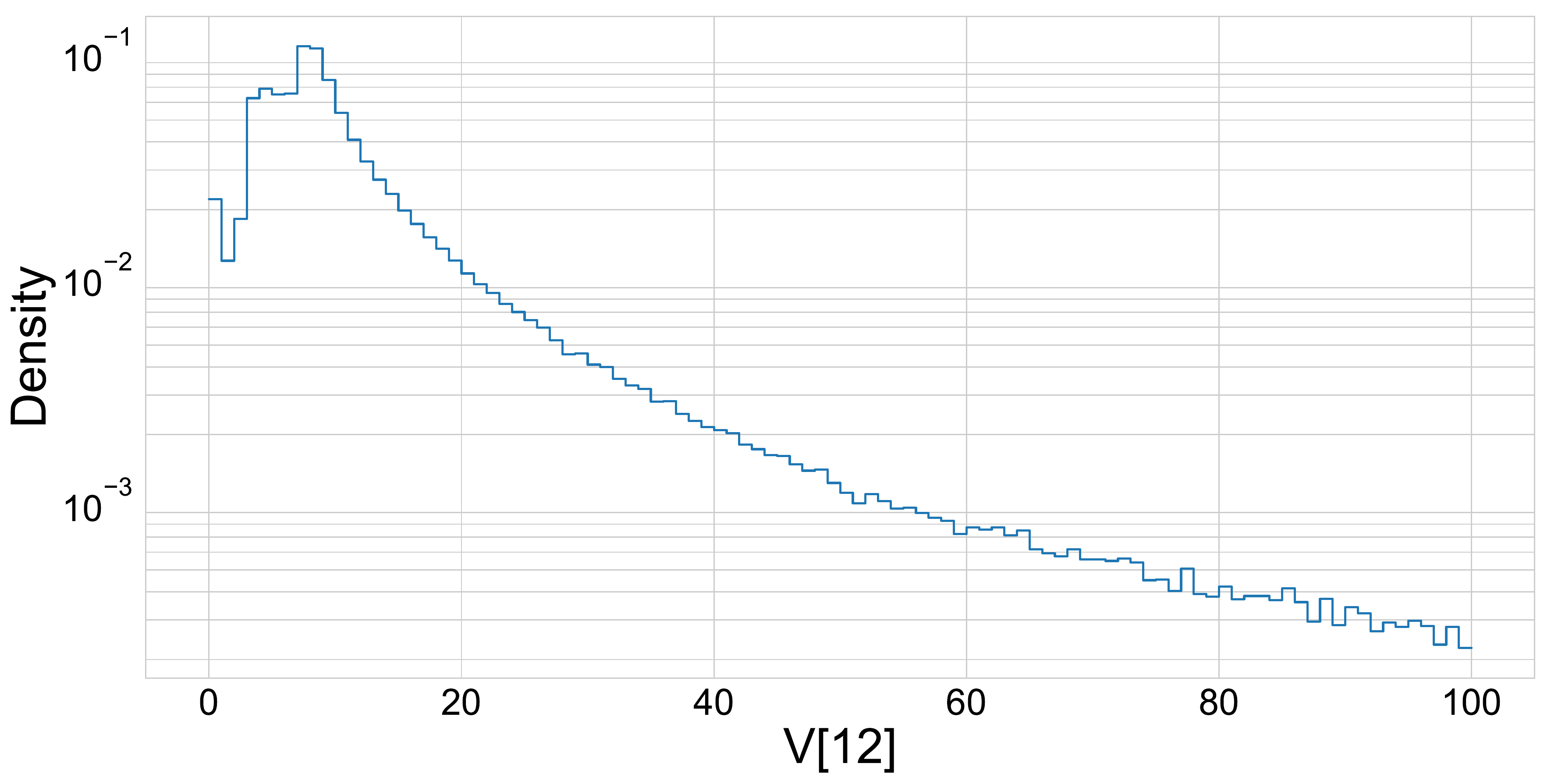}
            \end{center}
        \end{subfigure}
        \begin{subfigure}[t]{0.45\textwidth}
            \begin{center}
                \includegraphics[width=\textwidth]{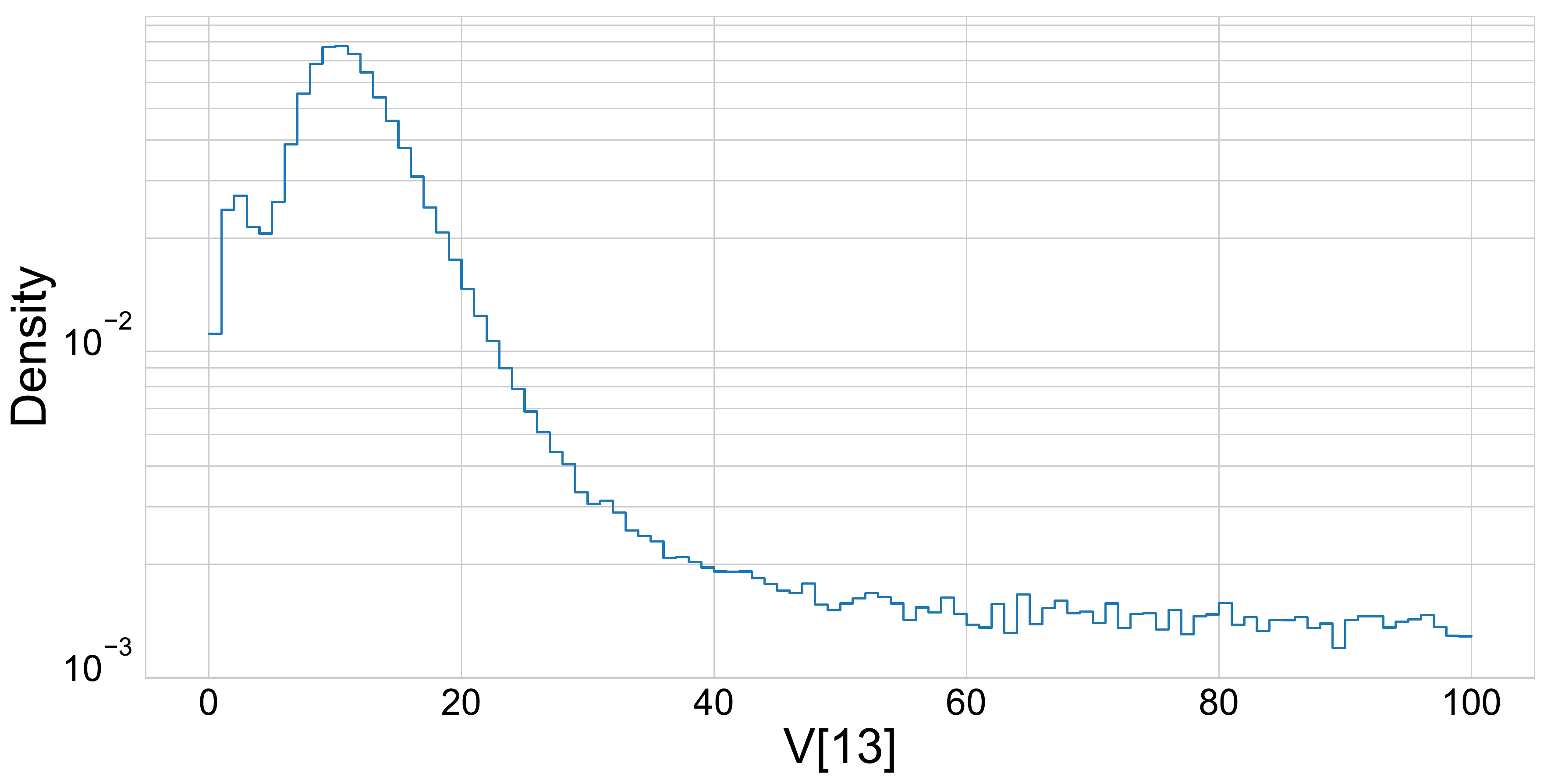}
            \end{center}
        \end{subfigure}
        \begin{subfigure}[t]{0.45\textwidth}
            \begin{center}
                \includegraphics[width=\textwidth]{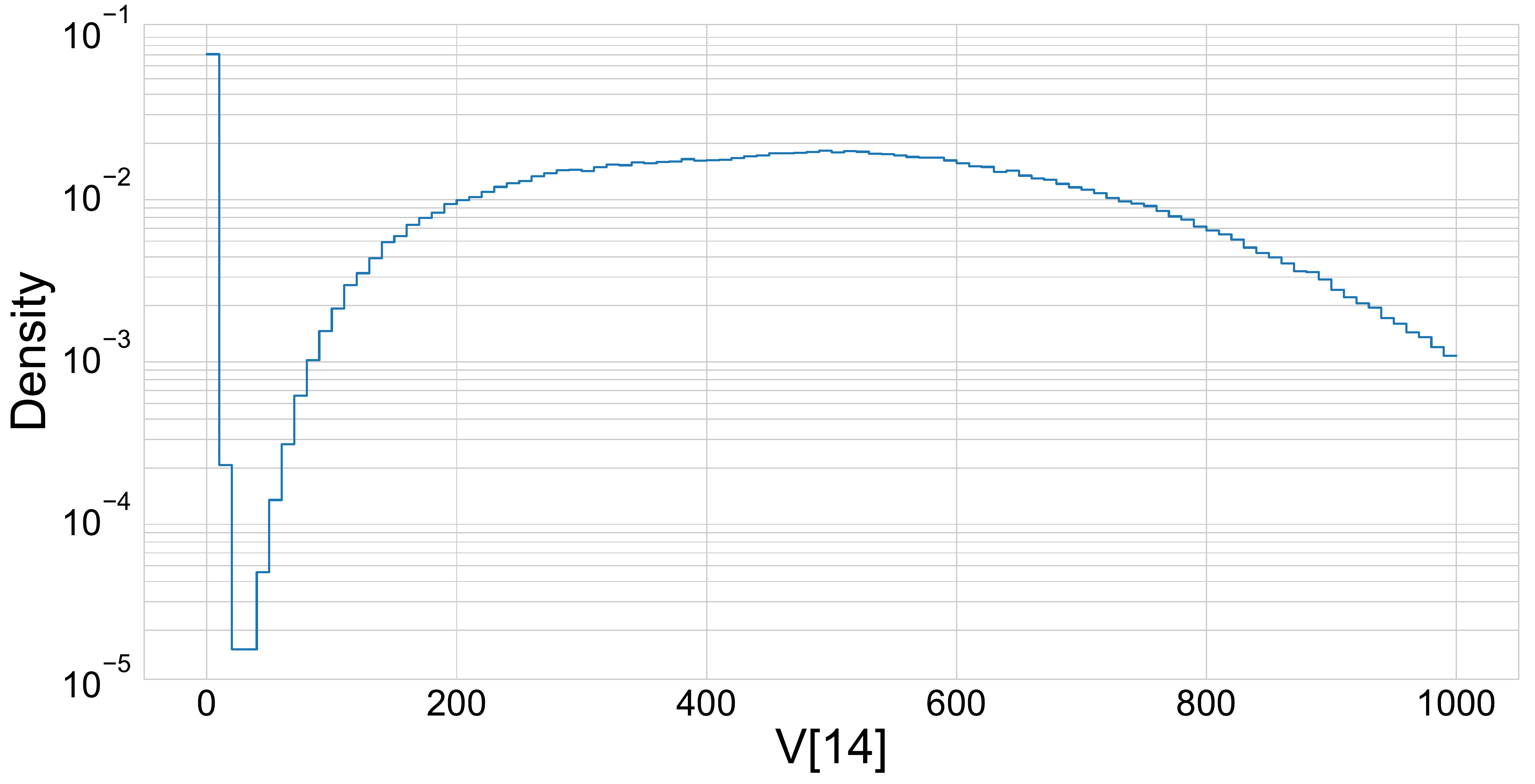}
            \end{center}
        \end{subfigure}
        \begin{subfigure}[t]{0.45\textwidth}
            \begin{center}
                \includegraphics[width=\textwidth]{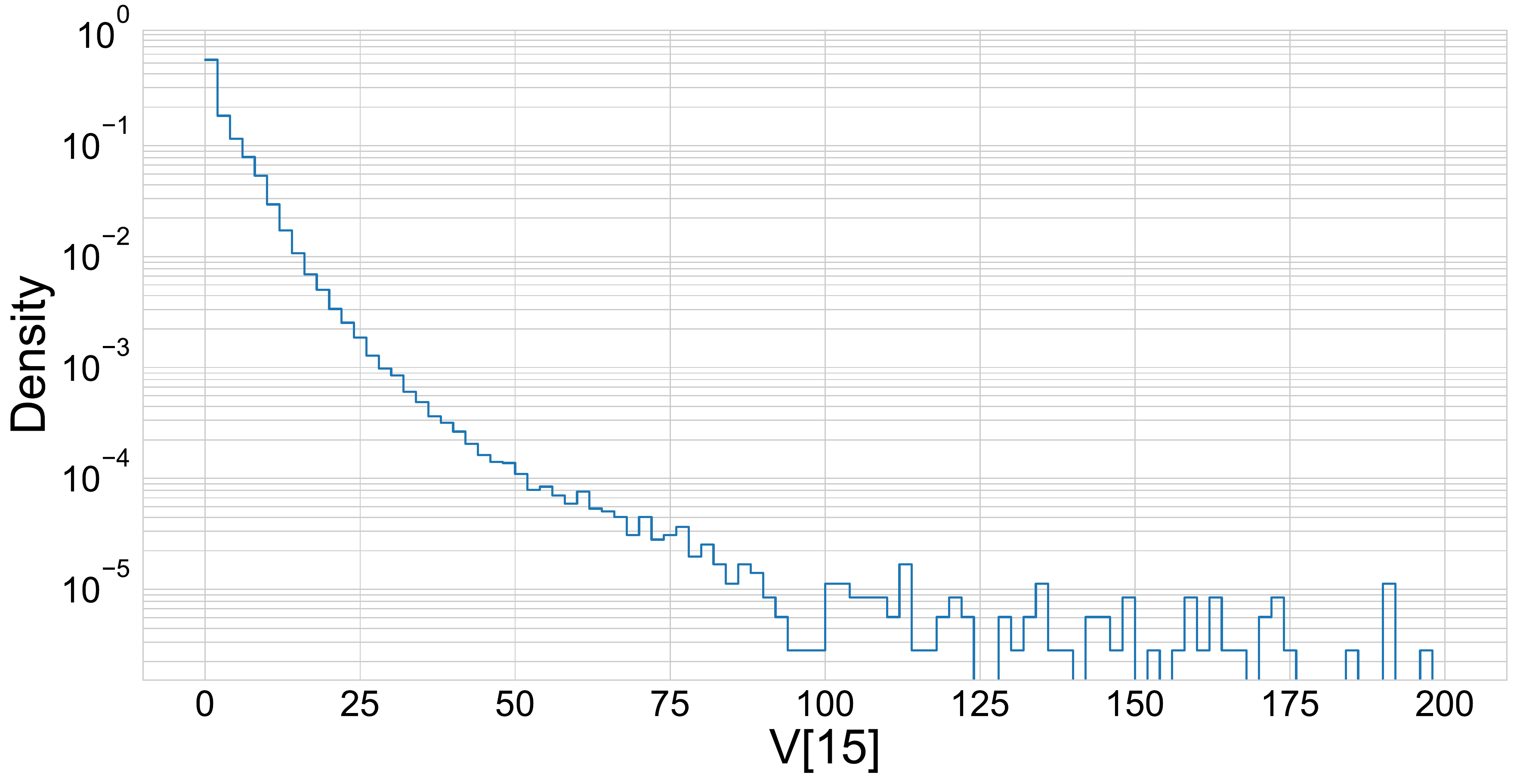}
            \end{center}
        \end{subfigure}
        \caption{\em Marginals of the six ``cluster" features. Top row: V[10],V[11]; second row: V[12],v[13]; bottom row: V[14],V[15]. Features are defined in Section~\ref{s:cluster}.}
        \label{f:cluster_feats}
    \end{center}
\end{figure}

\begin{figure}[ht]
    \begin{center}
        \begin{subfigure}[t]{0.30\textwidth}
            \begin{center}
                \includegraphics[width=\textwidth]{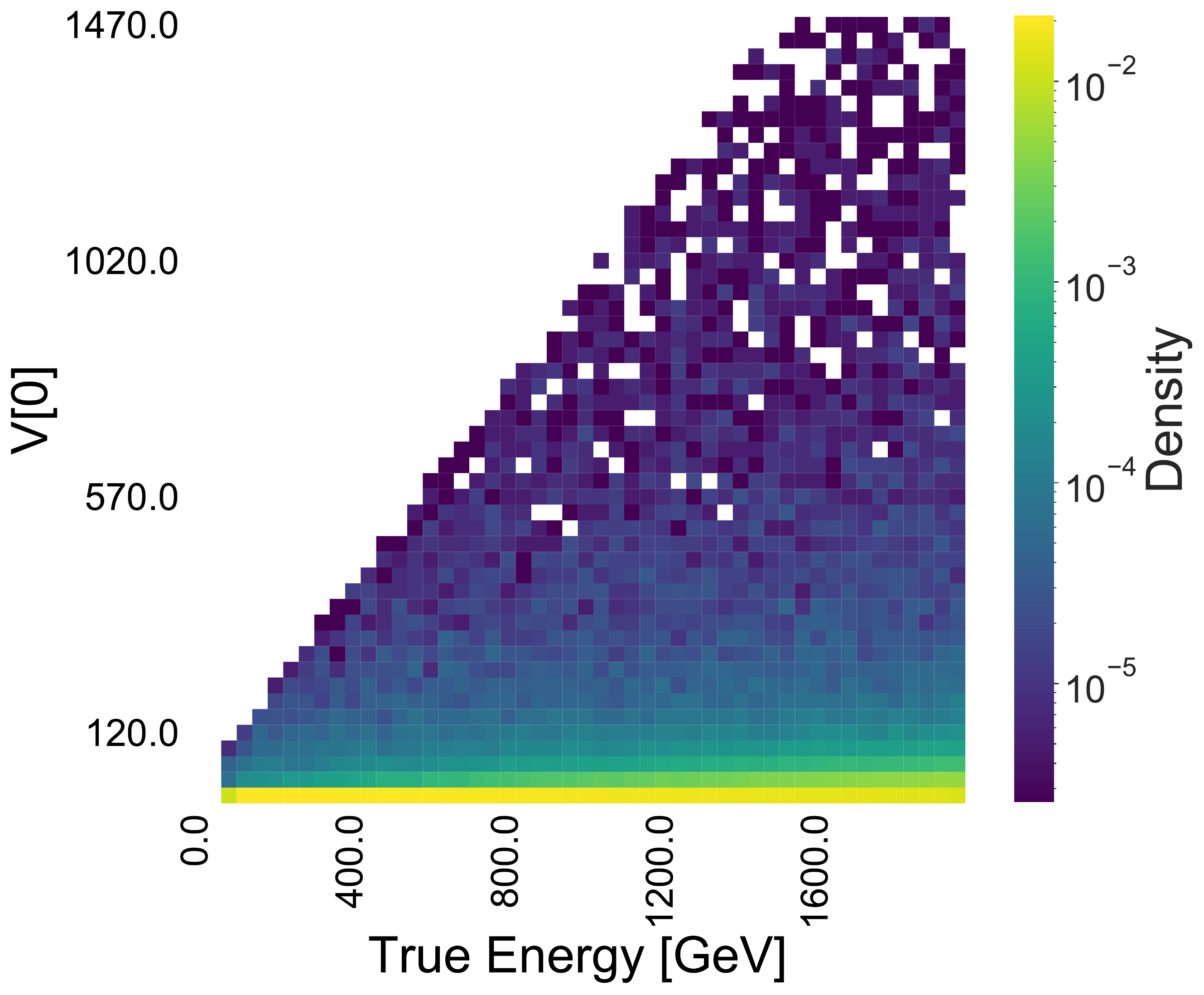}
            \end{center}
        \end{subfigure}
        \begin{subfigure}[t]{0.30\textwidth}
            \begin{center}
                \includegraphics[width=\textwidth]{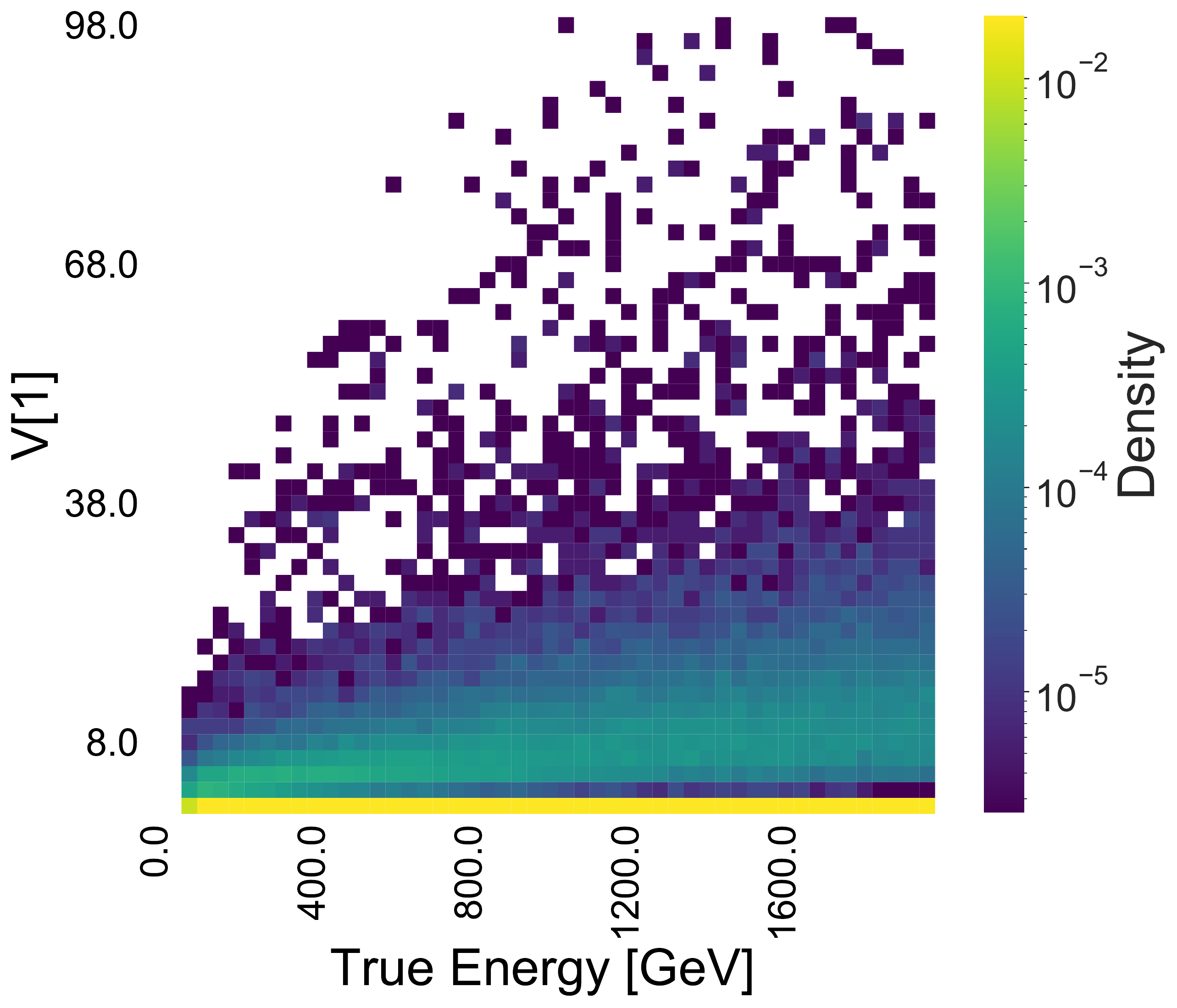}
            \end{center}
        \end{subfigure}
        \begin{subfigure}[t]{0.30\textwidth}
            \begin{center}
                \includegraphics[width=\textwidth]{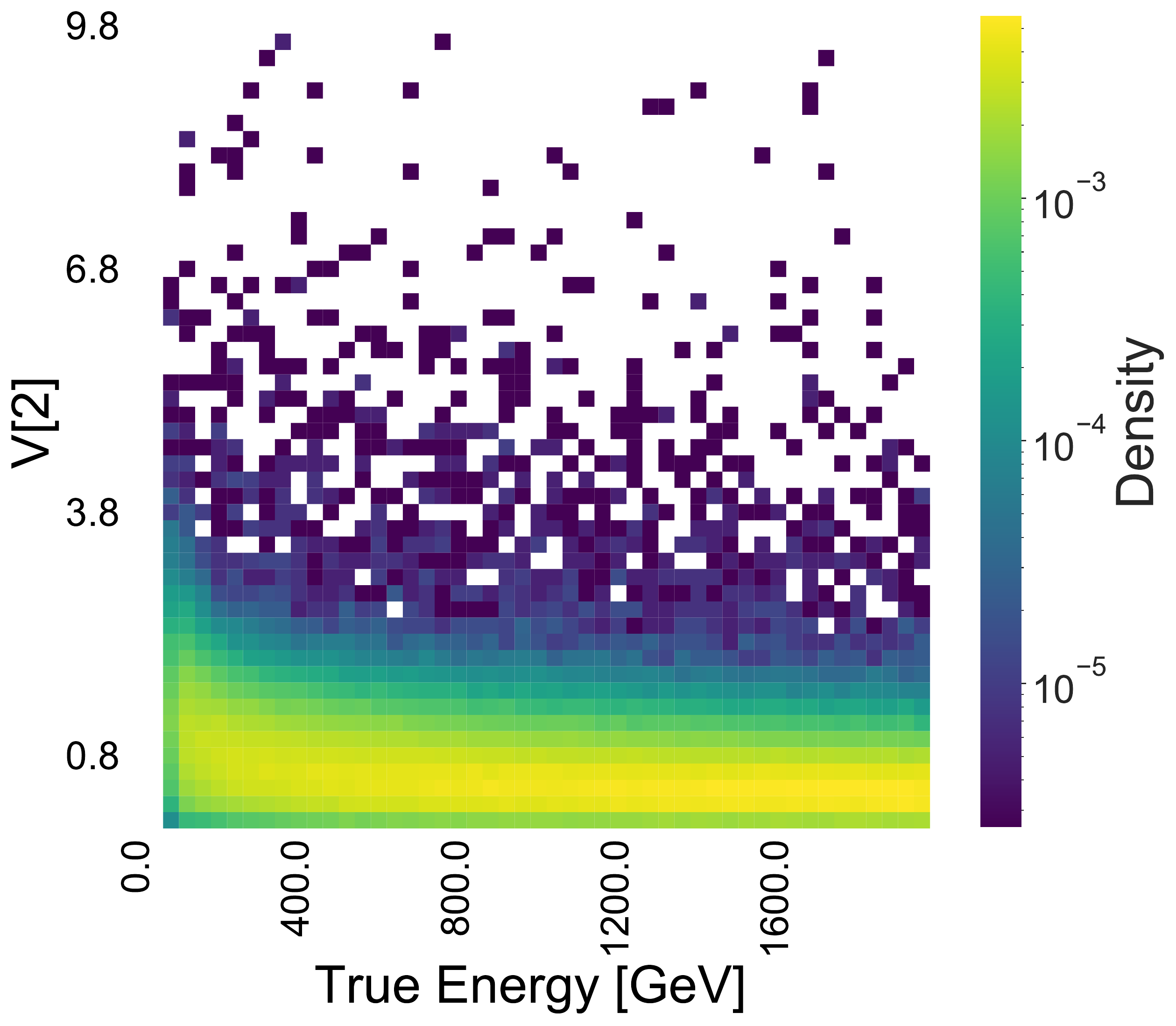}
            \end{center}
        \end{subfigure}
        \begin{subfigure}[t]{0.30\textwidth}
            \begin{center}
                \includegraphics[width=\textwidth]{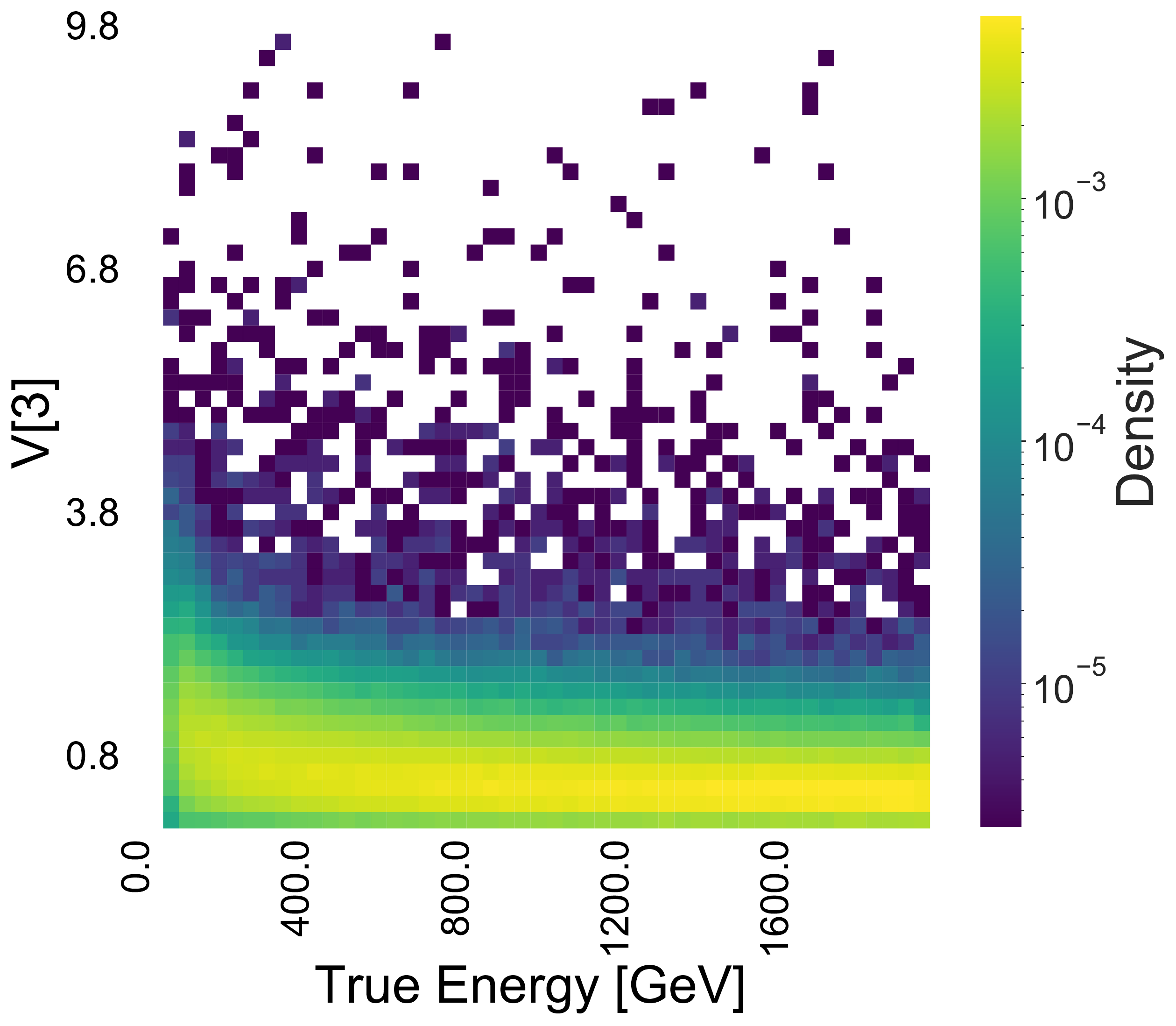}
            \end{center}
        \end{subfigure}\begin{subfigure}[t]{0.30\textwidth}
            \begin{center}
                \includegraphics[width=\textwidth]{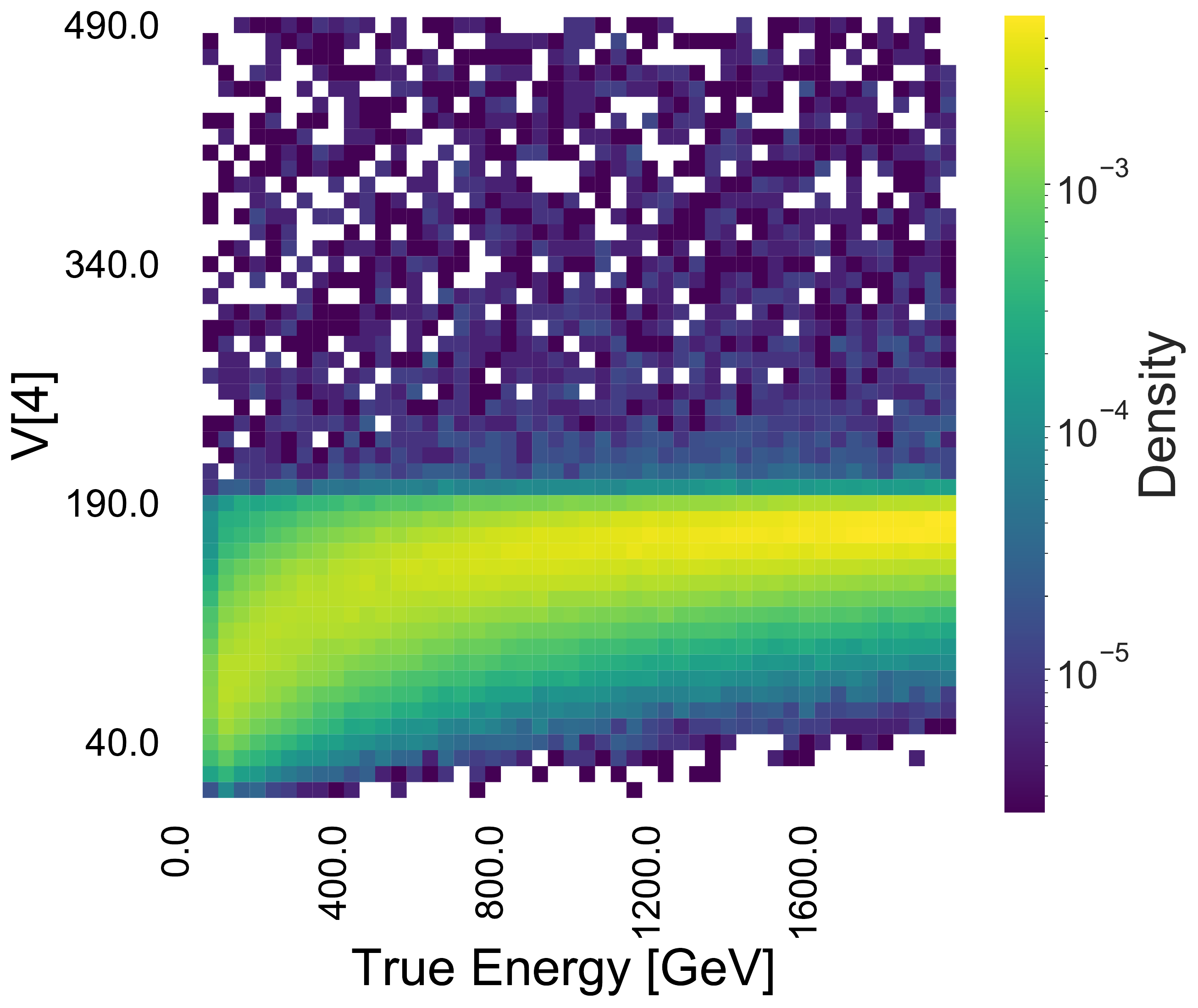}
            \end{center}
        \end{subfigure}\begin{subfigure}[t]{0.30\textwidth}
            \begin{center}
                \includegraphics[width=\textwidth]{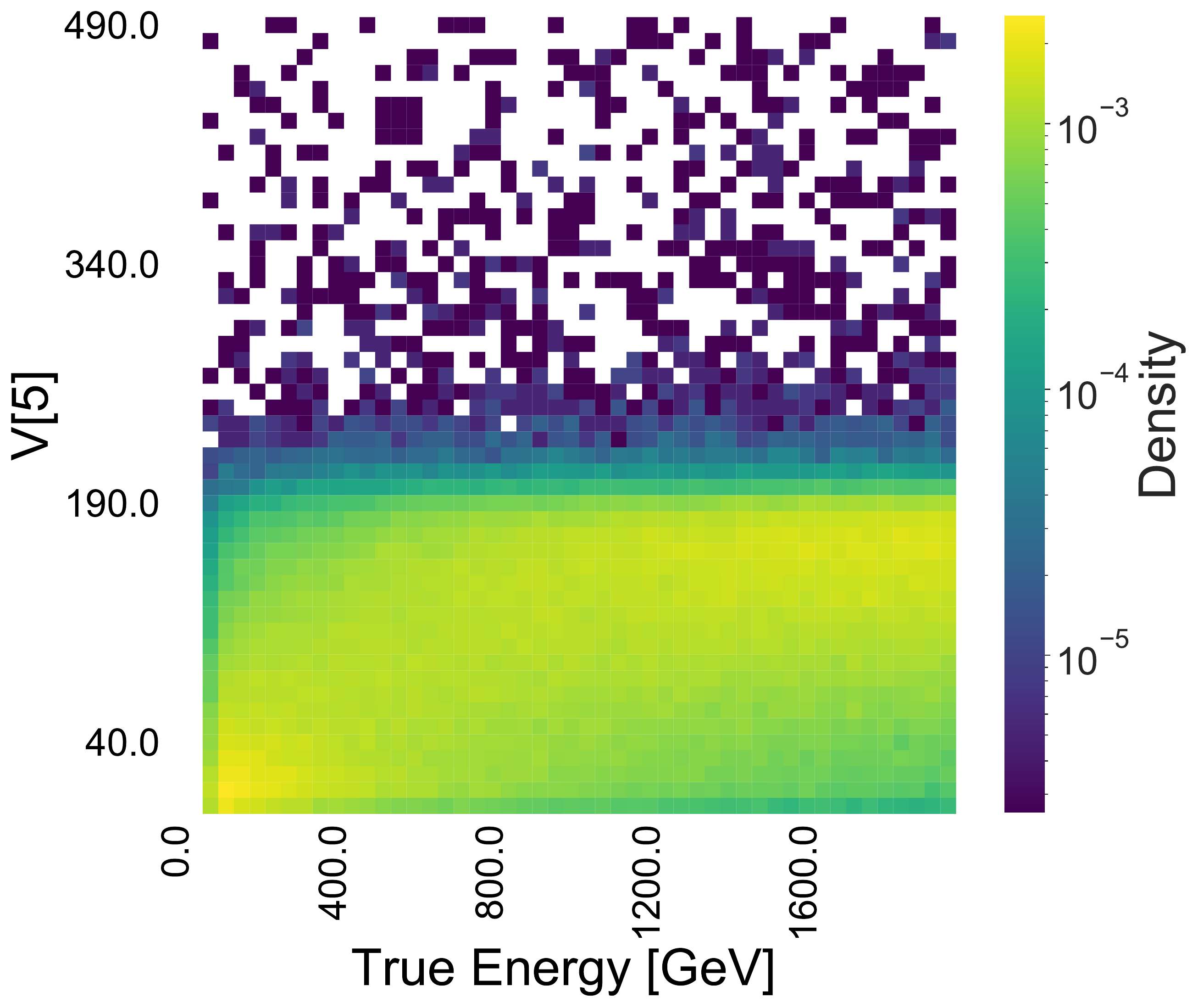}
            \end{center}
        \end{subfigure}
        \begin{subfigure}[t]{0.30\textwidth}
            \begin{center}
                \includegraphics[width=\textwidth]{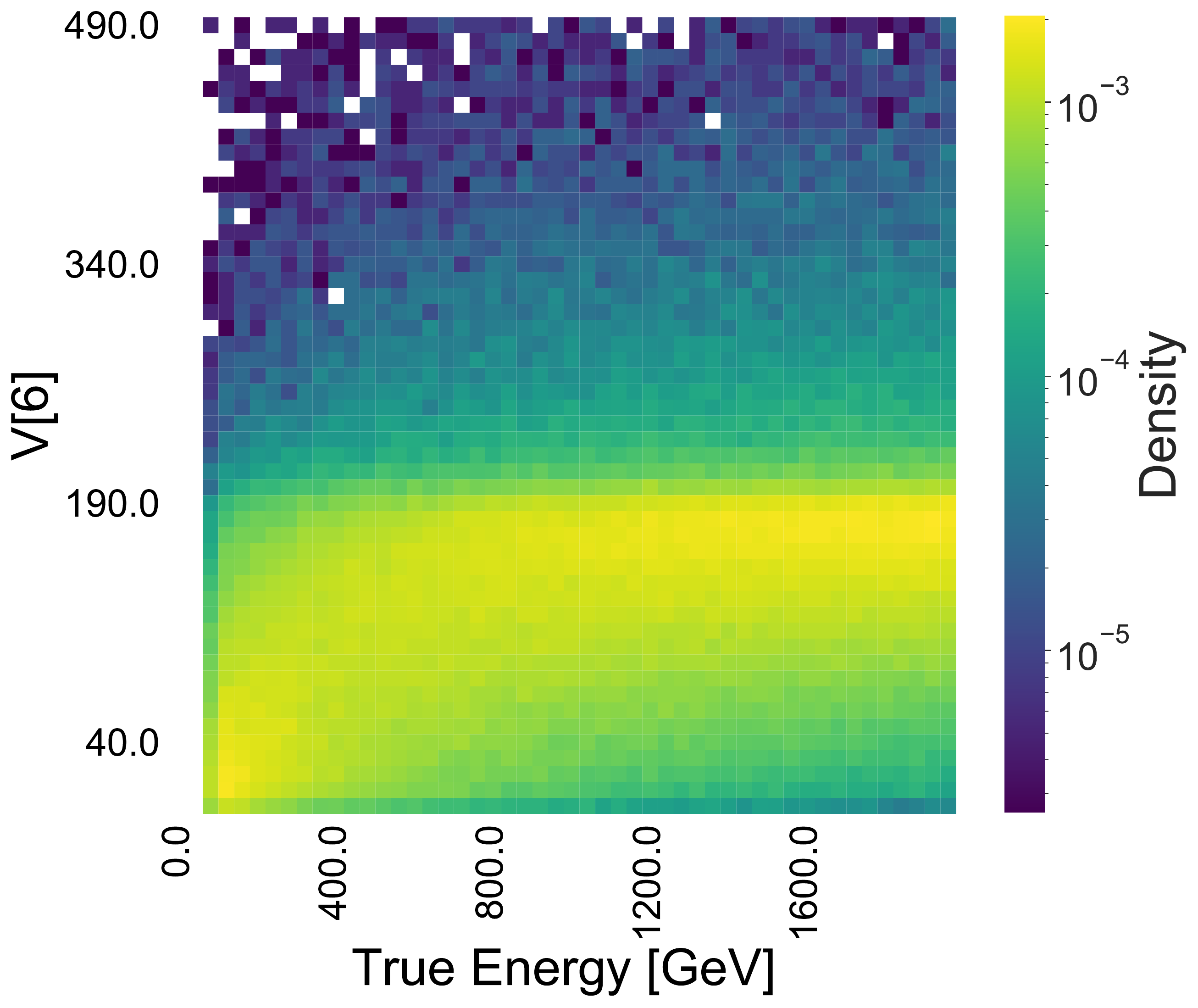}
            \end{center}
        \end{subfigure}
        \begin{subfigure}[t]{0.30\textwidth}
            \begin{center}
                \includegraphics[width=\textwidth]{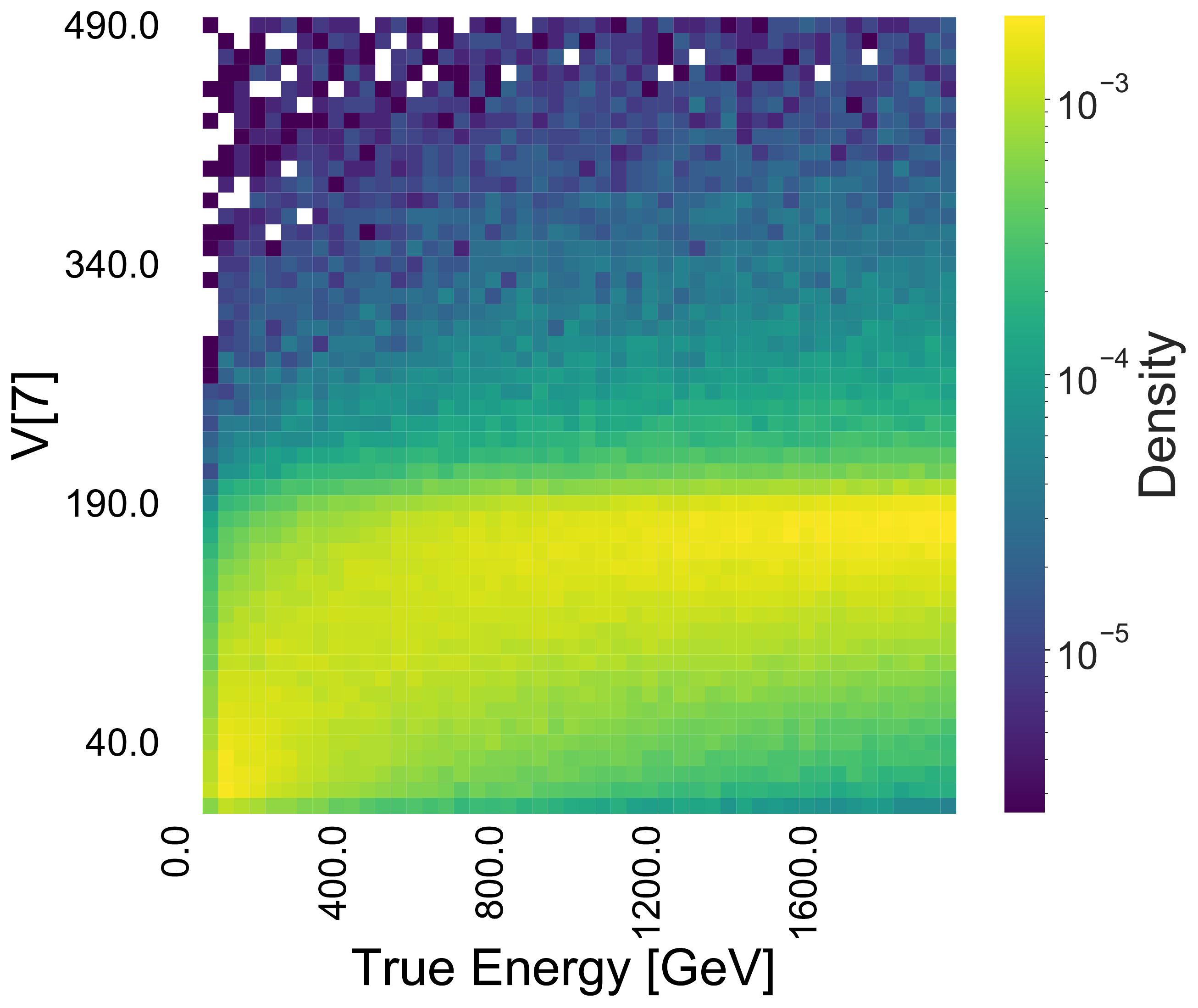}
            \end{center}
        \end{subfigure}
        \begin{subfigure}[t]{0.30\textwidth}
            \begin{center}
                \includegraphics[width=\textwidth]{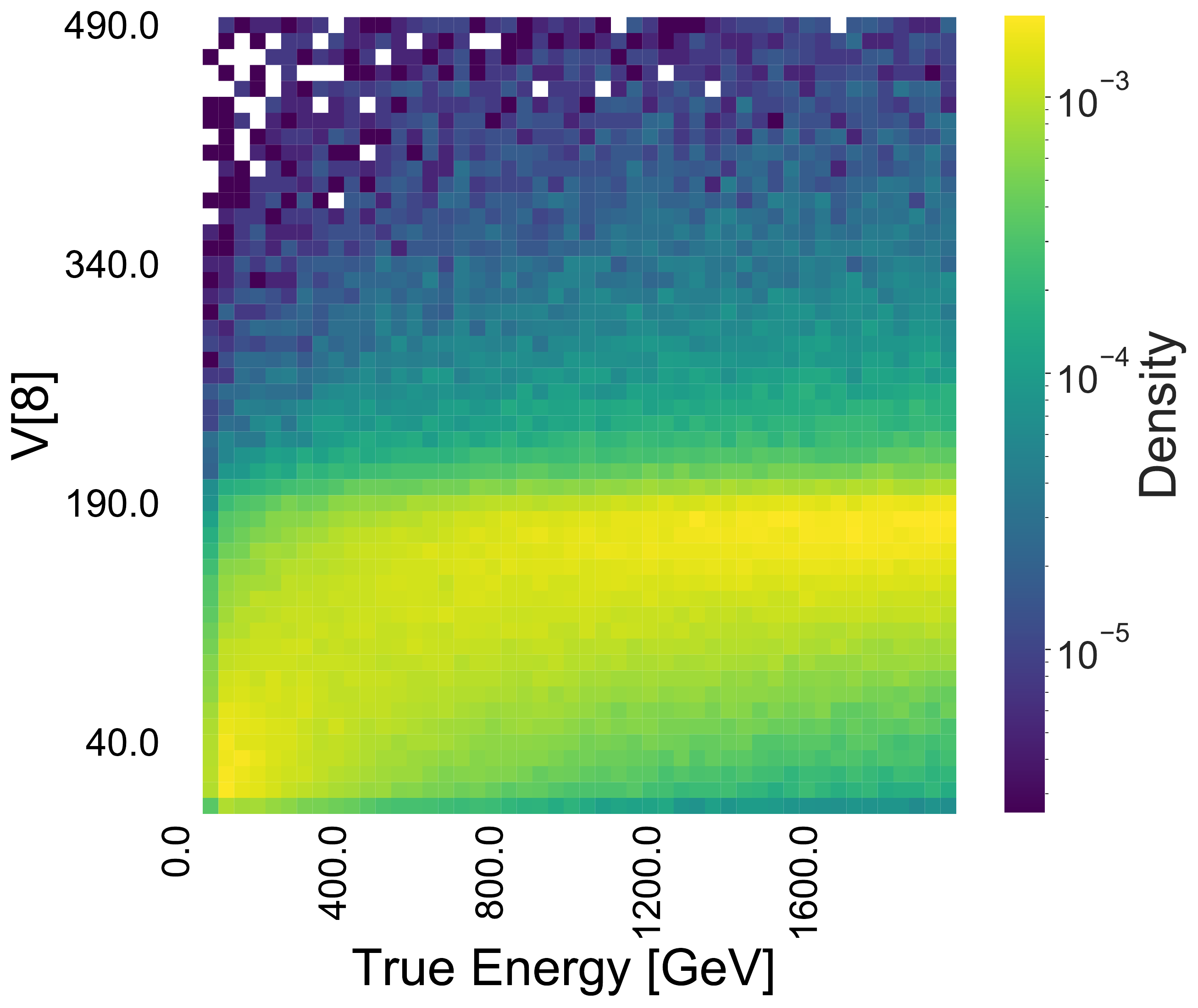}
            \end{center}
        \end{subfigure}
        \begin{subfigure}[t]{0.30\textwidth}
            \begin{center}
                \includegraphics[width=\textwidth]{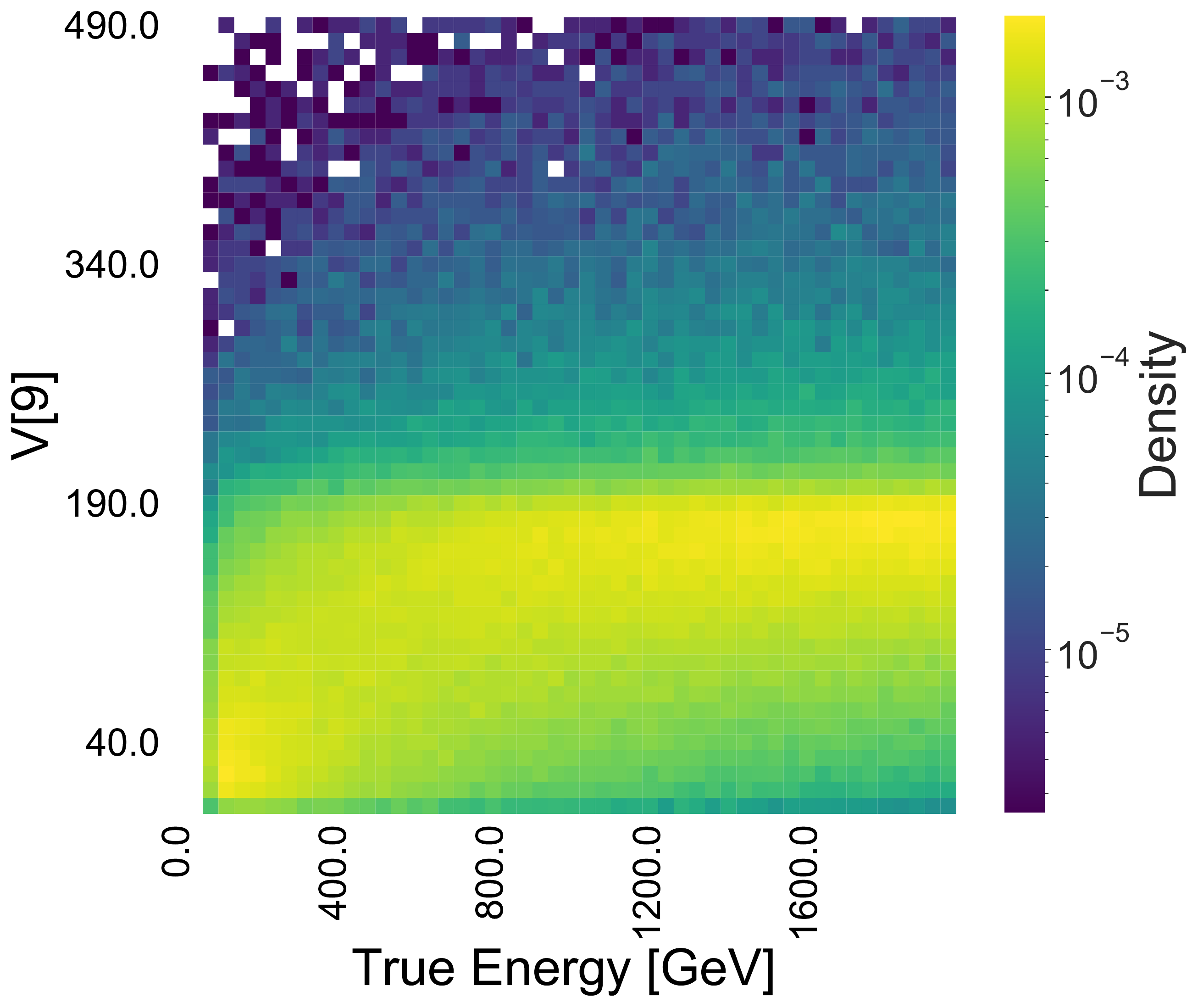}
            \end{center}
        \end{subfigure}
        \caption{\em 2D histograms showing the dependence of the 10 non-cluster-related event features (on the y axes) on true muon energy (on the x axes). Top row: V[0],V[1],v[2]; second row: V[3],V[4],V[5]; third row: V[6],V[7],V[8]; bottom row: V[9]. Features are defined in Section~\ref{s:other_features}.}
        \label{f:other_feats_2d}
    \end{center}
\end{figure}

\begin{figure}[ht]
    \begin{center}
        \begin{subfigure}[t]{0.45\textwidth}
            \begin{center}
                \includegraphics[width=\textwidth]{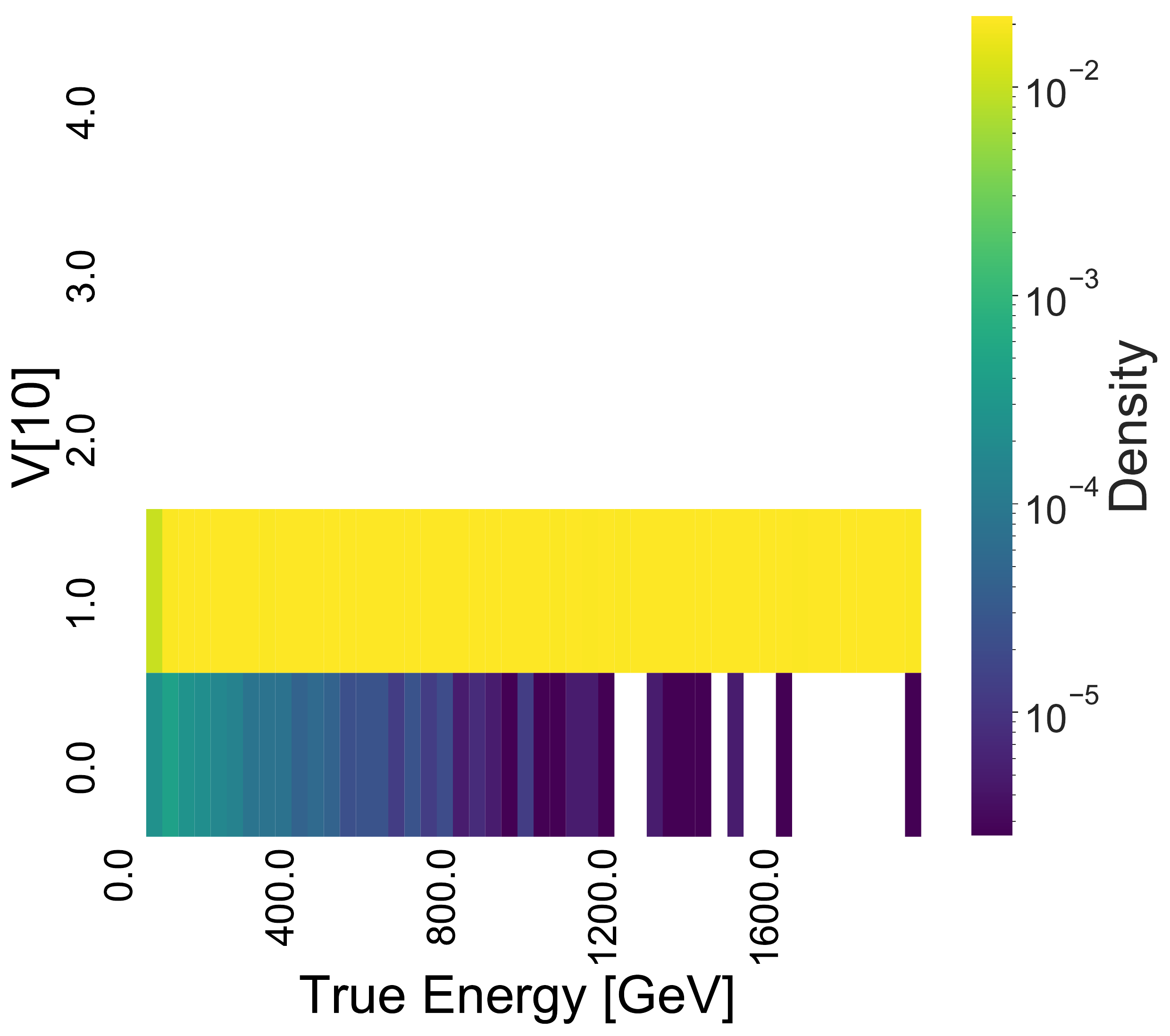}
            \end{center}
        \end{subfigure}
        \begin{subfigure}[t]{0.45\textwidth}
            \begin{center}
                \includegraphics[width=\textwidth]{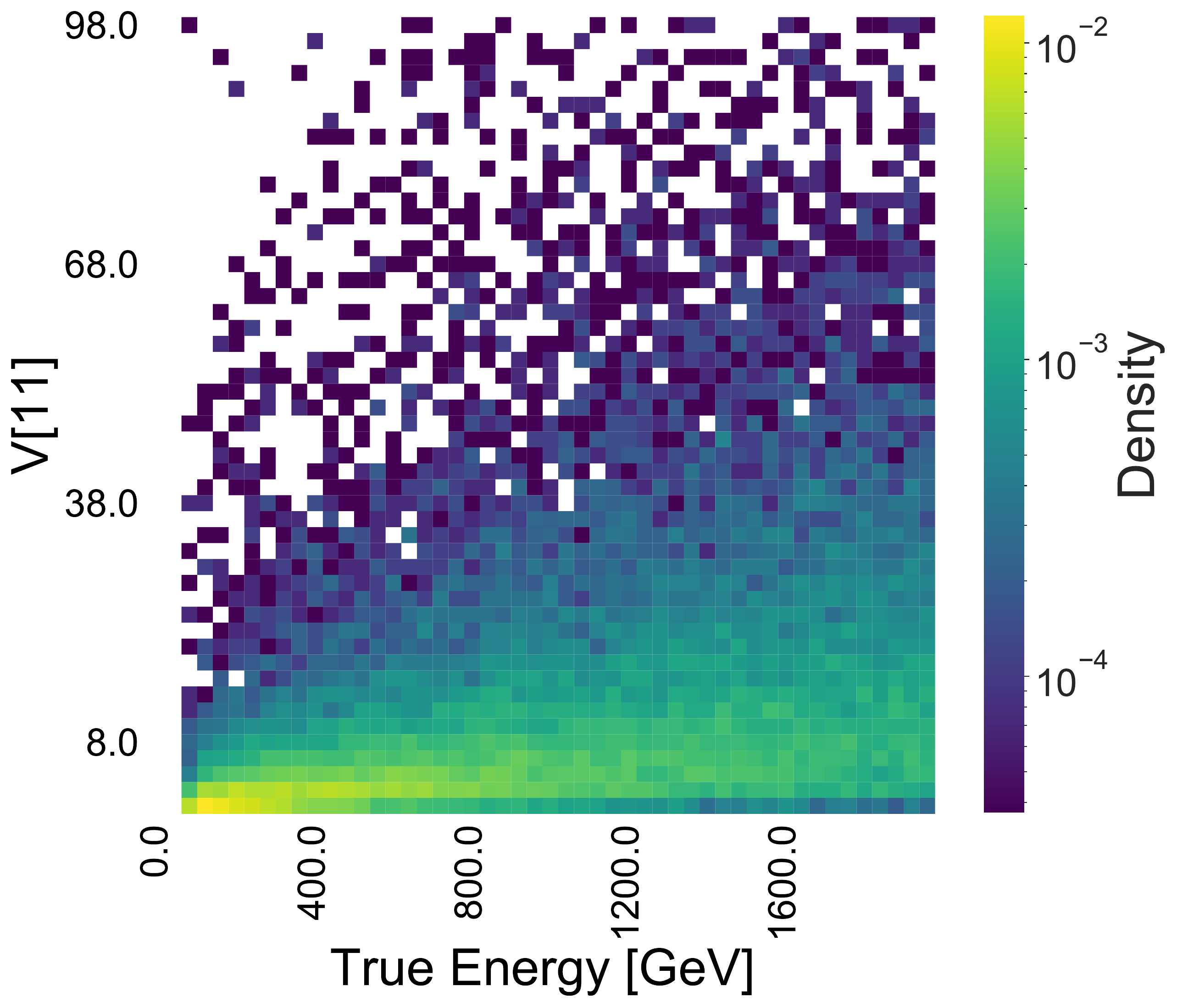}
            \end{center}
        \end{subfigure}
        \begin{subfigure}[t]{0.45\textwidth}
            \begin{center}
                \includegraphics[width=\textwidth]{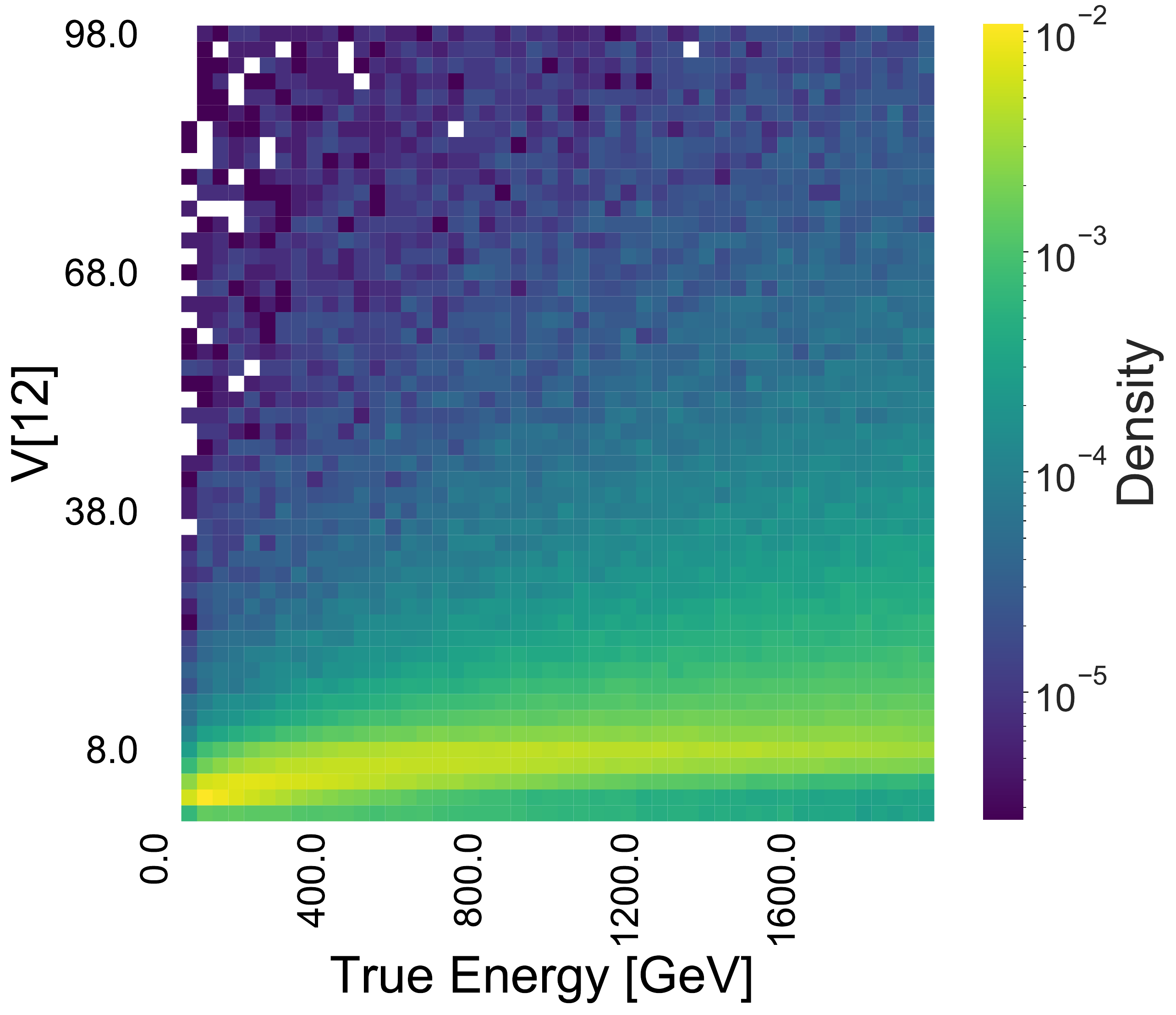}
            \end{center}
        \end{subfigure}
        \begin{subfigure}[t]{0.45\textwidth}
            \begin{center}
                \includegraphics[width=\textwidth]{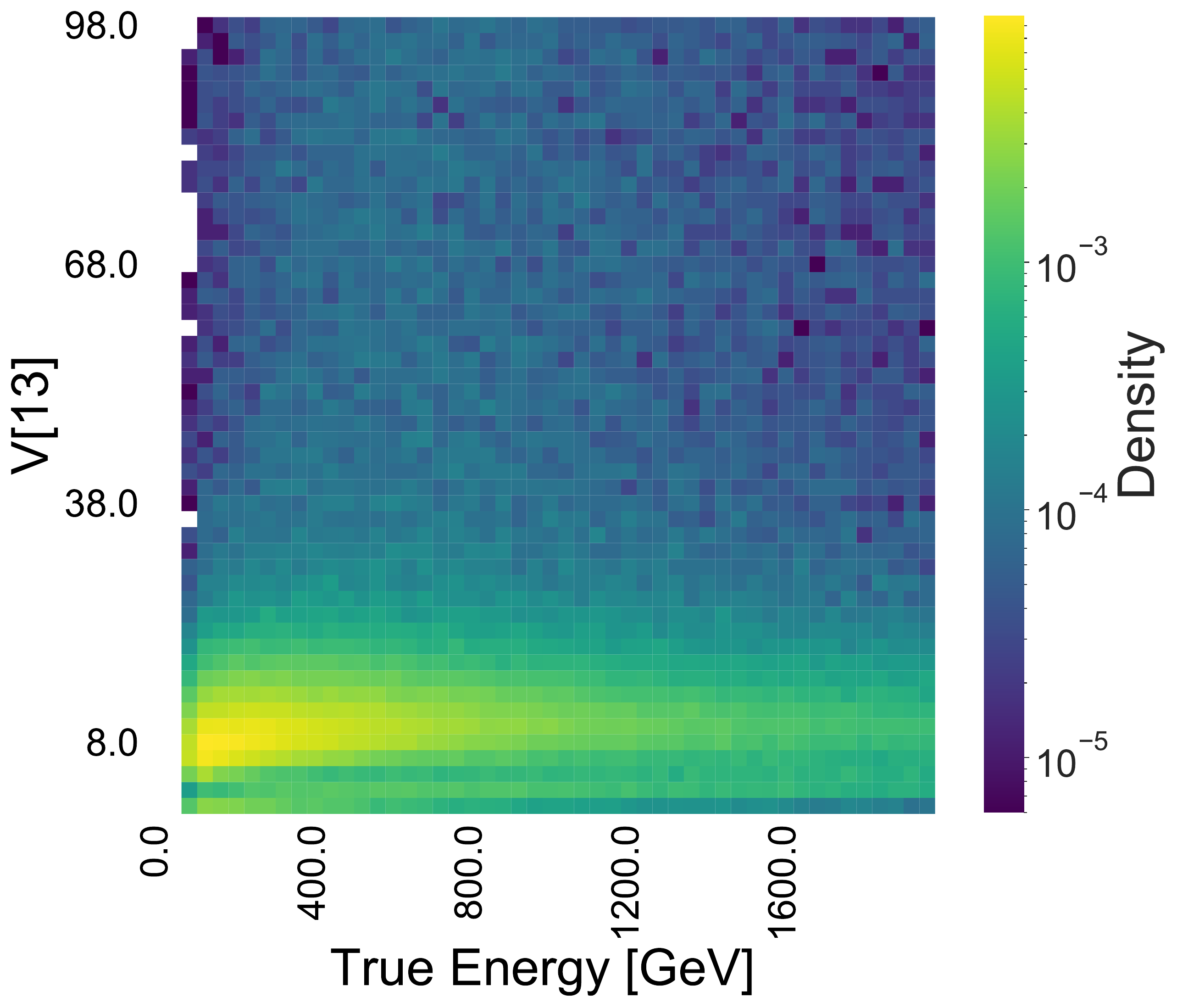}
            \end{center}
        \end{subfigure}
        \begin{subfigure}[t]{0.45\textwidth}
            \begin{center}
                \includegraphics[width=\textwidth]{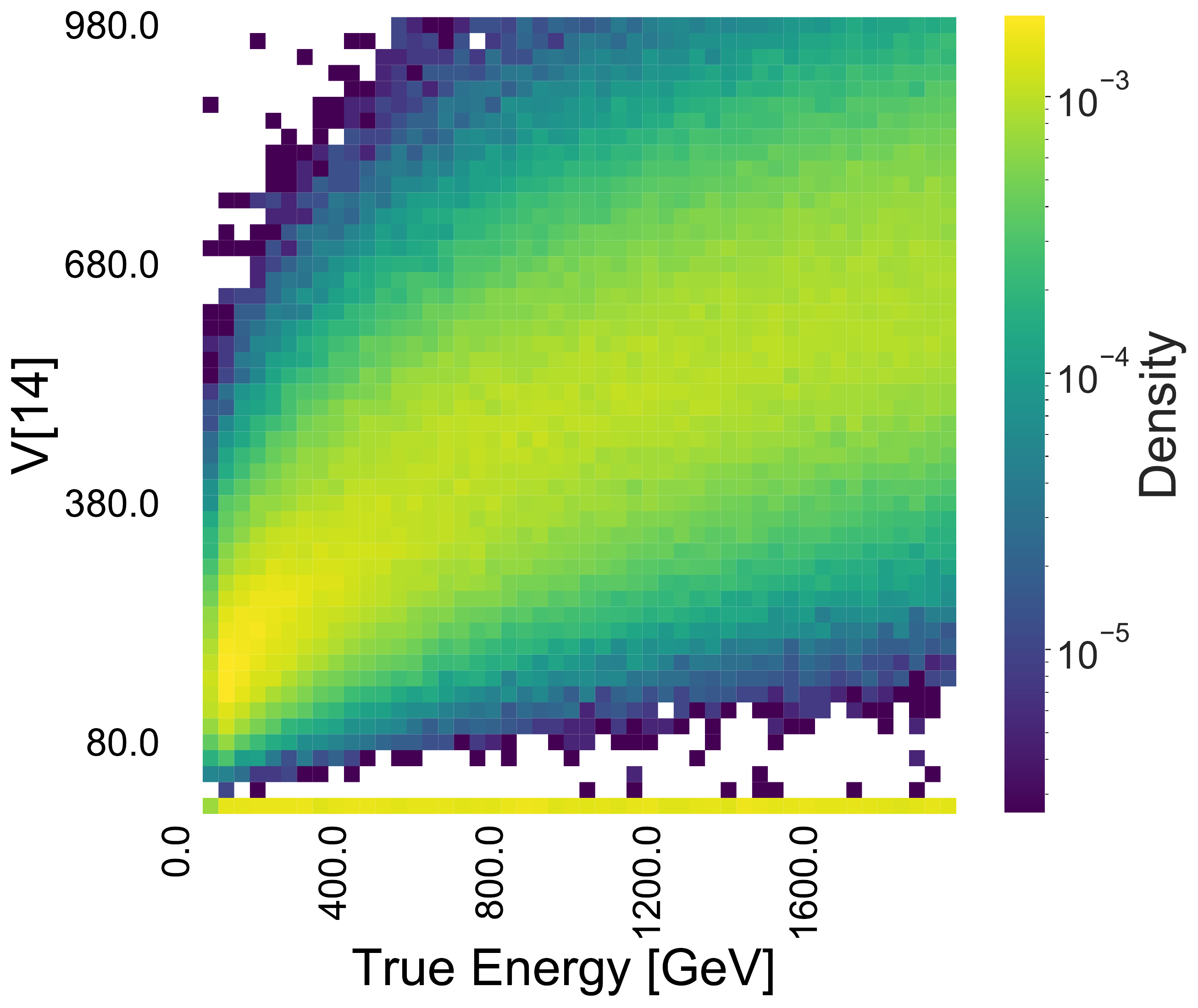}
            \end{center}
        \end{subfigure}
        \begin{subfigure}[t]{0.45\textwidth}
            \begin{center}
                \includegraphics[width=\textwidth]{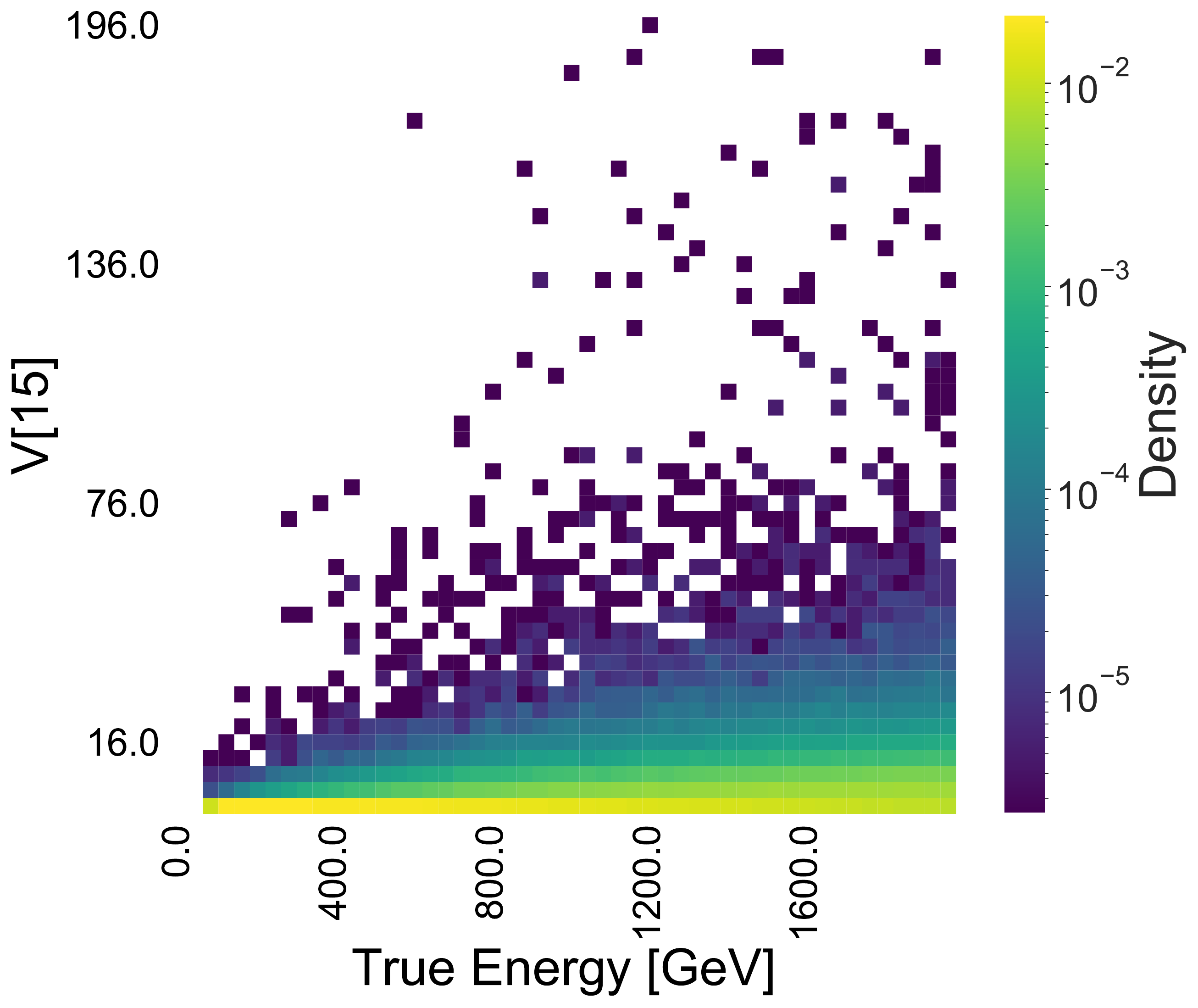}
            \end{center}
        \end{subfigure}
        \caption{\em 2D histograms showing the dependence of six cluster-related features (on the y axes) on true muon energy (on the x axes). Top row: V[10],V[11]; second row: V[12],v[13]; bottom row: V[14],V[15]. Features are defined in Section~\ref{s:cluster}.}
        \label{f:cluster_feats_2d}
    \end{center}
\end{figure}

\clearpage

%%%%%%%%%%%%%%%%%%%%%%%
\section{The regressor}\label{s:regressor}

While the problem of learning the patterns of soft radiation from high-energy muons would lend itself optimally to a study involving deep learning tools, we choose to employ for our regression task a customized \kNN algorithm\footnote {For a general introduction to \kNN algorithms see \cite{knn}.}, essentially based on the ``Hyperball" algorithm developed for CMS searches and used for a search of Supersymmetric Higgs bosons in LHC Run 1 data\cite{hbb}. The result of this technique allows for a direct comparison of the extractable information with the non-spatial part of the detector output, namely the total deposited energy, as the regression using that single input may be performed in exactly the same manner as the more complex regression using the full set of variables.

Our custom version of the \kNN algorithm uses feature sub-sampling \cite{featsample} as a source of stochasticity to build an ensemble of weak learners, whose relative importance (parametrized by weights adding up to 1) is learnt via gradient boosting \cite{boost,boost2} to a calorimeter-inspired loss function. In the following section the algorithm is described in detail.

\subsection{The algorithm}\label{s:algorithm}

Simulated data are processed to extract the 16 features of each event along with the generator-level information on muon energy. Marginals of the 16 features are shown in \autoref{f:other_feats} and \autoref{f:cluster_feats}, and 2D histograms of the dependence of their value on the true muon energy are shown in \autoref{f:other_feats_2d} and \autoref{f:cluster_feats_2d}. The event features of \num{396000} training events simulated with a flat distribution of muon true-energy are read in by a custom $c^{++}$ algorithm that performs the \kNN-driven regression. 

First, some hyperparameters are defined: \par
\begin{itemize}
    \item $K_0$ and $K_1$ the numbers of neighbors used by the algorithm in two stages (see below);
    \item a learning rate is also defined to decide the width of the gradient descent steps; a multiplier $\lambda=0.004$ of the partial derivative of the loss with respect to the individual features has been used for the results shown below; the learning rate is increased during minimization whenever the direction of steepest descent remains the same, and is reset to an exponentially decreasing value when the maximum gradient changes direction.
    \item $N_{batch}$ is the number of events considered during the optimization cycle.
\end{itemize}

\noindent
Other parameters are described below. The algorithm works as follows:\par

\begin{enumerate}
    \item Input data are standardized by subtracting means $\hat{V_n}$ and dividing each of the $n=1...16$ features by their all-sample root-mean-square, $RMS_n$, of the corresponding marginal distribution, so that $V_n=(V_n-\hat{V_n})/RMS_n$.
    \item Data are divided into training, bias correction\footnote{The bias correction sample is required to obtain a better matching of true and predicted energy, as the \kNN algorithm is sensitive to bulk properties of the input data, and therefore tends to regress values toward the center of the input distribution (\SI{1000}{\gev}).} (bc), and test samples. These contain respectively up to $(N_{\text{train}}, N_{\text{bc}}, N_{\text{test}}) = (\num{396000},\num{3000},\num{12000})$ events. 
    \item An ensemble of $N_L$ weak regressors is created through feature sub-sampling by randomly selecting sets of flags, of dimension equal to the feature space (16 variables), which define the features used by each learner.
    So, {\em e.g.}, if a flag vector is $f=(1,1,1,1,0,0,0,0,0,0,0,0,0,0,0,0,0)$, the corresponding \kNN regressor works in the 4-dimensional subspace spanned by the first four variables. The number of regressors $N_L$ is chosen at run time, and so is the average number $N_{active}$ of flags set to 1 for each regressor.
    \item An optimization cycle is run using training data ($j=1....N_{train})$, to find the best value of weights $W_i$ ($i=1...N_L$) such that a linear combination of the predictions $P_{j,i}$ of the $N_L$ regressors, $P_j=\sum{W_i P_{j,i}}$ produces the lowest value of a loss function (the definition of the loss and the details of the optimization cycle are given {\em infra}); the $N_L$ weights are constrained to add up to one, but are independently allowed to be negative, adding flexibility to the estimators.
    \item Once weights are optimized, a cycle is run on the bias correction sample (of indices $j=1...N_{\text{bc}}$), in order to extract a correction function. Each event is regressed using optimized weights, and the distribution of the uncorrected predictions $P_j$ is fitted to an invertible function of the respective true values $T_j$, extracting the dependence $P=F(T)$. 
    \item The fitted function is inverted to $T=F^{-1}(P)$, and used to obtain corrected predictions $P_{j,\text{corr.}}=F^{-1}(P_j)$.
\end{enumerate}

\noindent 
We now discuss the optimization cycle of the weak regressors. Initially, weights $W_i$ are initialized to equal values, $W_i=1/N_L$. Then, within a cycle the following operations are performed:\par

\begin{enumerate}
    \item $N_{\text{batch}}$ events are chosen at random from the training sample (typical used values of $N_{\text{batch}}$ are in the 500-10000 events range), while preserving a uniform distribution in true muon energy ($N_{\text{batch}}/19$ events are required in each of 19 100-GeV-wide bins from 100 to 2000 GeV).
    \item A loop on $j=1...N_{\text{batch}}$ is executed:
    for each event $j$, and for each regressor $i$, another loop is performed on all training events not coincident to $j$; the $K_0$
    events closest to $j$ are identified. The distances are defined as $D_i^2=\sum_{n=1...16} (V_{nj}-V_{nm})^2$.
    \item The relative importance ($W_{HB,n}$) 
    of each of the active features in each tested point $j$ of feature space is determined by a separate calculation (described later in \autoref{s:HB}), involving the identified $K_0$ closest points to the considered event $j$.
    \item A new loop over all training events not coincident with event $j$ is performed to find the $K_1$ events closest to $j$, this time using the HB-improved metric $D_{HB,nj}^2 = \sum_{n=1...16} W_{HB,nj}^2*(V_{nj}-V_{nm})^2$.
    \item The uncorrected prediction on muon energy from weak regressor $i$  is then computed as $P_i=\sum_{m=1...K_1} T_m/K_1$.
    \item A linear fit is performed on the distribution of $P_i$ as a function of $T_i$ for the considered $N_{\text{batch}}$ events, extracting a correction function $P_i=F(T_i)$; the correction function is inverted and used to obtain corrected predicted values $P_{i,\text{corr.}}=F^{-1}(P_i)$.
    \item The $N_L$ corrected predictions (one per each learner) are combined in an ensemble prediction $P=\sum_{i=1...N_L} W_i P_i$, and used to calculate a batch loss inspired by calorimetric measurement resolution, 
    $L = \sum_{i=1...N_{\text{batch}}} (T-P)^2/T$.
    \item The loss is derived with respect to each regressor weight $W_i$; the largest derivative is used to determine a step in a stochastic gradient descent cycle, which updates the corresponding weight in search for a minimum of L. The cycle is repeated until the loss reaches a minimum value.
\end{enumerate}

\noindent
At  the end of the above cycle, which should sample a sufficient number of batches to allow for a convergence, the optimized weights $W_i$ may be used for the final test run.

A bias-correction cycle can then be operated, which only evaluates the unbiased prediction using weights $W_i$ and then fits to the full distribution of $P$ versus $T$, to obtain functional parameters of the correction function. Finally, a cycle is performed on all $N_{\text{test}}$ test events. Each prediction is corrected for the inverse dependence derived on the $N_{\text{bc}}$ bias correction sample, obtaining the final prediction for each test event.

\subsection {Hyperball optimization \label{s:HB}}

There remains to explain how weights $W_{HB}$ are extracted from the $K_0$ closest training events to each batch event $j$. If we consider the fact that we are trying to extract a correction from a multi-dimensional space whose features $V$ possess complex interdependence and dependence with the true muon energy $T$, we see that the averaging procedure produced by a \kNN algorithm might benefit from exploiting that structure. The Hyperball algorithm attempts to gauge the local properties of the dependence of features on target, by directly estimating the bias incurred by averaging the target over a finite interval of each feature.

\begin{center}
\begin{figure}[h!]
\centerline{\includegraphics[width=0.6\textwidth]{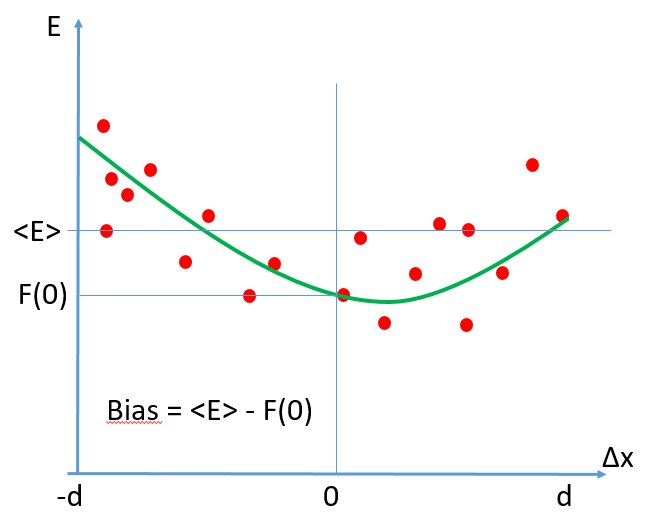}}
\caption{\em Graphical description of the bias in the calculation of the energy E from the averaging of events located at distances $\Delta x$ along one variable, if the true dependence of $E$ on $x$ is quadratic.}
\label{f:HB}
\end{figure}
\end{center}

\noindent
If we consider a one-dimensional interval of relative distances $[-d,d]$ on a feature $V$ for event $j$, and a set of $j$'s $K_0$ closest neighbors, we may picture the regression to $T_j$ as the averaging of the true energy $T$ of the neighbors. This procedure works as long as the dependence of $T$ on $V$ is linear in the interval. If the dependence is more complex, a bias will be present due to having integrated the complex function $T(V)$ in a finite interval to estimate the value at the center of the interval. An estimate of the bias for each feature is produced by comparing the average of $T(V)$ in the hyperball range $[-d,d]$ with the value of the offset parameter $c$ of a quadratic function $f(\Delta V)=a \Delta V^2 + b \Delta V + c$ interpolating the observed values of $T$ as a function of their distance along $V$ from $V_j$ (see \autoref{f:HB}).

The calculation, while of course subjected to statistical imprecisions as well as systematic ones (arising, {\em e.g.}, from the non-quadratic behavior of the true dependence), allows one to derive separate estimates of the biases $W_{HB_nj}$ for each of the $n$ active directions around each test point $j$ of the feature space. They may then be used to ``reshape" the hyperellipsoid that allows one to find the most relevant $K_1$ closest training events to the event for which a prediction is required.

\section{Tests and results}\label{s:tests}

\subsection{Choice of algorithm parameters}

The algorithm described in \autoref{s:algorithm} requires significant CPU time to run an optimization over a reasonable number of batch cycles. A compromise must therefore be struck on the choice of parameters. 

 The first parameter to consider is number of events in each hyperball $K_1$ used for the averaging of the muon energies. In general one would desire this parameter to be large enough that a precise average of energies is obtained; however, there is a trade-off to consider with the variability ofmuon energy on the event features. In addition, with 400k events, setting $K_1$ to more than {\em e.g.} 40 events would mean to partition the data into less than O(10k) independent regions. This would place a tight bound on the number of active dimensions in which to define the hyperballs: for instance, $N_{\text{active}}=5$ would mean that each direction of space would be partitioned in only $x=(10000)^{1/5} \simeq 6$ really independent regions, while $N_{\text{active}}=6$ would already be definitely too large, yielding only four independent regions per feature. While the above is a back-of-the-envelope estimate, it guides our choice of $K_1$ and $N_{active}$.
 
 We start investigating how to optimize the hyperparameters by targeting $N_{\text{active}}=4-8$ and investigating the range $K_1=5-50$ for our further studies. Those ranges should allow the algorithm to exploit the properties of the feature space sufficiently well. The next choice is the number of weak regressors. Given the large number of possible combinations of few active dimensions in a set of 16, and the intrinsic capability of the stochastic gradient loop to exploit the best linear combination of regressors, one might expect --and indeed we have found-- that there is no upper limit on their number to improve the performances; however, improvements are very small when $N_L$ grows to be larger than 20 or so.
 What limits $N_L$ is the large computing demand posed by the nested loops of the algorithm described in the previous section. We have limited our investigations to $N_L$ up to 25, and indeed we present here results of that choice.
 
 The batch size should be large enough to allow for a meaningful bias correction of the regressor outputs before they are linearly combined. This parameter is also CPU-limited, and we have typically used values in the 200-2000 event range for it. 
 
 Finally, the number $K_0$ of events in the larger initial hyperballs used to extract $W_{HB}$ values should be larger than $K_1$, lest no meaningful value can be associated with the extracted biases in each active dimension. We typically set $K_0 = 4 K_1$ in our training runs.
 
 \subsection {Results \label{s:results}}

Rather than discussing all the combinations of hyperparameters we have probed in the search for an optimal working point, we describe here only the results of a good compromise between the various choices on which we have settled. For the results presented in this section, the parameters and hyperparameters of the \kNN regressor are the following:\par
\begin{itemize}
    \item $K_0=10$ events are used for the hyperballs determining the $W_{HB}$ weights; this number is probably smaller than it could be usefully set to, but the computing time required for larger values makes it a reasonable practical choice.
    \item $K_1=5$ events are used for the hyperballs extracting the predictions;
    \item $N_L=25$ \kNN regressors are combined linearly, with as many weights $W$ obtained by an optimization cycle ran on batches of $N_{\text{batch}}=500$ events and 100 loops of stochastic gradient descent per batch;
    \item a small number $N_{\text{active}}=3.2$ average active dimensions is used for the construction of the 25 regressors. We found that besides keeping the computing time down, that choice exploits better the individual information of the event features, avoiding the ``regression to the mean" of the predicted energy caused by the large correlations existing among them.
\end{itemize}

\noindent 
Because of the imperfect optimization of the hyperparameters quoted above, the results of this Section are not the final answer on the performance of the described algorithm on the task at hand, as a global hyperparameters optimization has not been attempted yet. In addition, longer training runs than those so far performed will probably further improve the regression performance. On the other hand, results are already indicative of the kind of solution that the algorithm will give to the problem.

One very important question that we originally set out to address is whether meaningful information on muon energy can be extracted from the spatial distribution of the energy depositions. Indeed, if an electron can perfectly well be measured by the proportionality of yielded light in a single scintillating block, why is it necessary to aim for a finely segmented detector for muons? Our preconceived answer to that question is that muons behave in an entirely different way from electrons. While for electrons energy deposition is the result of a number of fractioning processes that terminates at the reaching of a (low) critical threshold, which makes the proportionality of total light yield to deposited energy a strong summary statistic, for muons it is rather the effect of multiple soft photon radiations along the particle path in the detector material; and radiation probability is a strong function of the muon energy itself, as is emission angle. Given the above difference it appears reasonable to expect that useful information for muons may be extracted from a high-spatial resolution pattern of those energy depositions. 

We tested the hypothesis that spatial information helps the regression task by comparing our regression attempts, which indeed leverage very detailed, although summarized, information on the spatial radiation pattern, to a regression based on the simple correlation of the total deposited energy ($E_{\text{sum}}$) with the muon incoming energy. The latter is extracted by running the same \kNN algorithm, with the same general settings, on a single regressor with only the first flag (the one corresponding to $V[0]$) set on. A bias correction is applied to the prediction as a function of true energy, as in the case of the combined \kNN prediction discussed above. We refer to this model as the ``energy-sum model".

\begin{center}
\begin{figure}[h!]
\centerline{\includegraphics[width=0.9\textwidth]{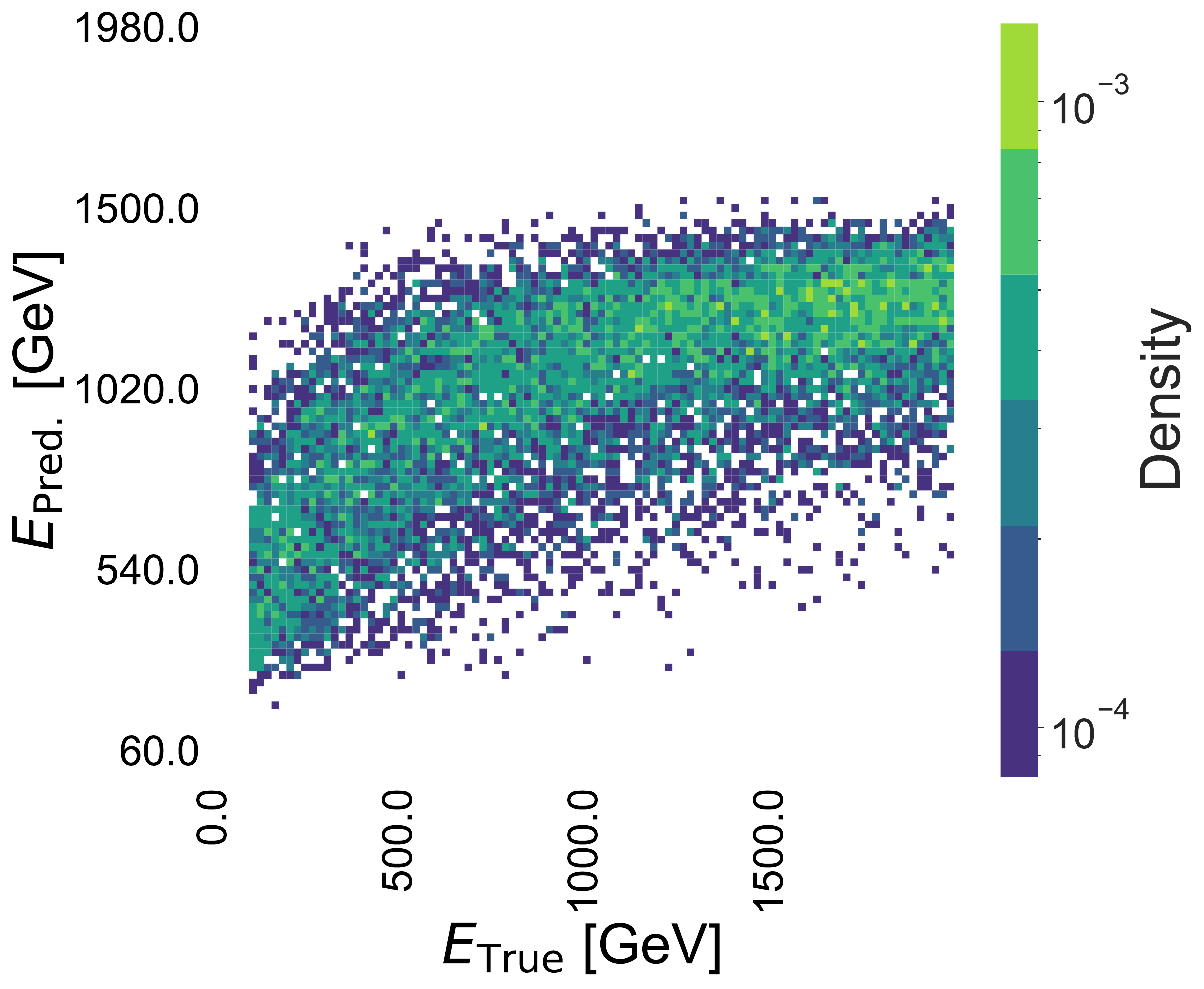}} %uncorrectedkNN.PNG}}
\caption{\em 2D histogram of uncorrected \kNN prediction versus true energy for test data. A bias towards the mean of the interval of energies of training data is evident, and it is the object of a correction that is applied by inverting the linear fit to the distribution, as described in the text.}
\label{f:uncorrectedkNN}
\end{figure}
\end{center}

\begin{figure}[ht]
    \begin{center}
        \begin{subfigure}[t]{0.45\textwidth}
            \begin{center}
                \includegraphics[width=\textwidth]{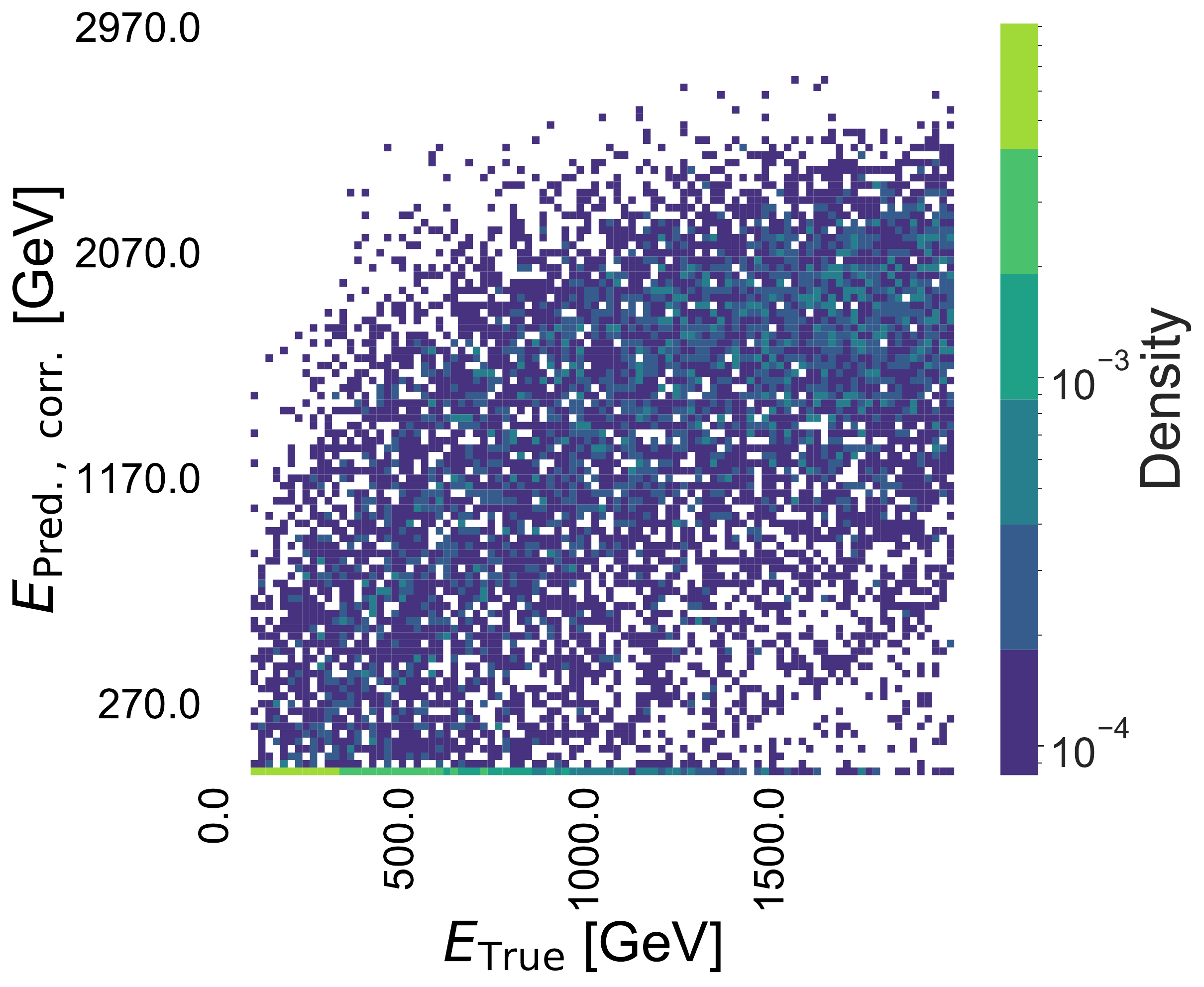}
            \end{center}
        \end{subfigure}
        \begin{subfigure}[t]{0.45\textwidth}
            \begin{center}
                \includegraphics[width=\textwidth]{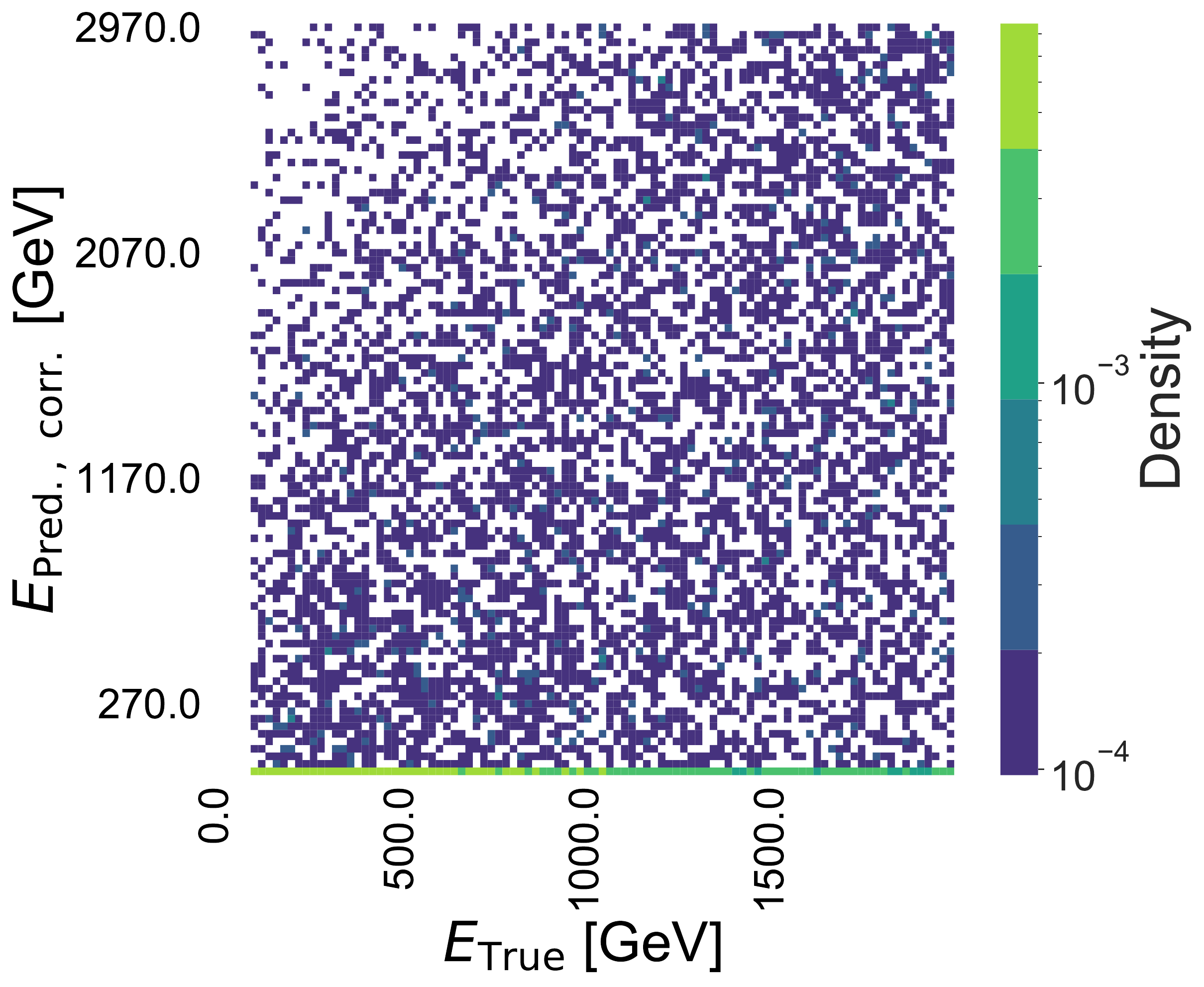}
            \end{center}
        \end{subfigure}
        \caption{\em Left panel:
2D histogram of corrected \kNN prediction versus true energy for test data; right panel: same, for the corrected prediction of the energy sum model. A profile of the distributions is overlaid in red to both graphs to show the average value of the prediction in bins of true muon energy.}
        \label{f:correctedkNN}
    \end{center}
\end{figure}

\begin{center}
\begin{figure}[h!]
\centerline{\includegraphics[width=0.9\textwidth]{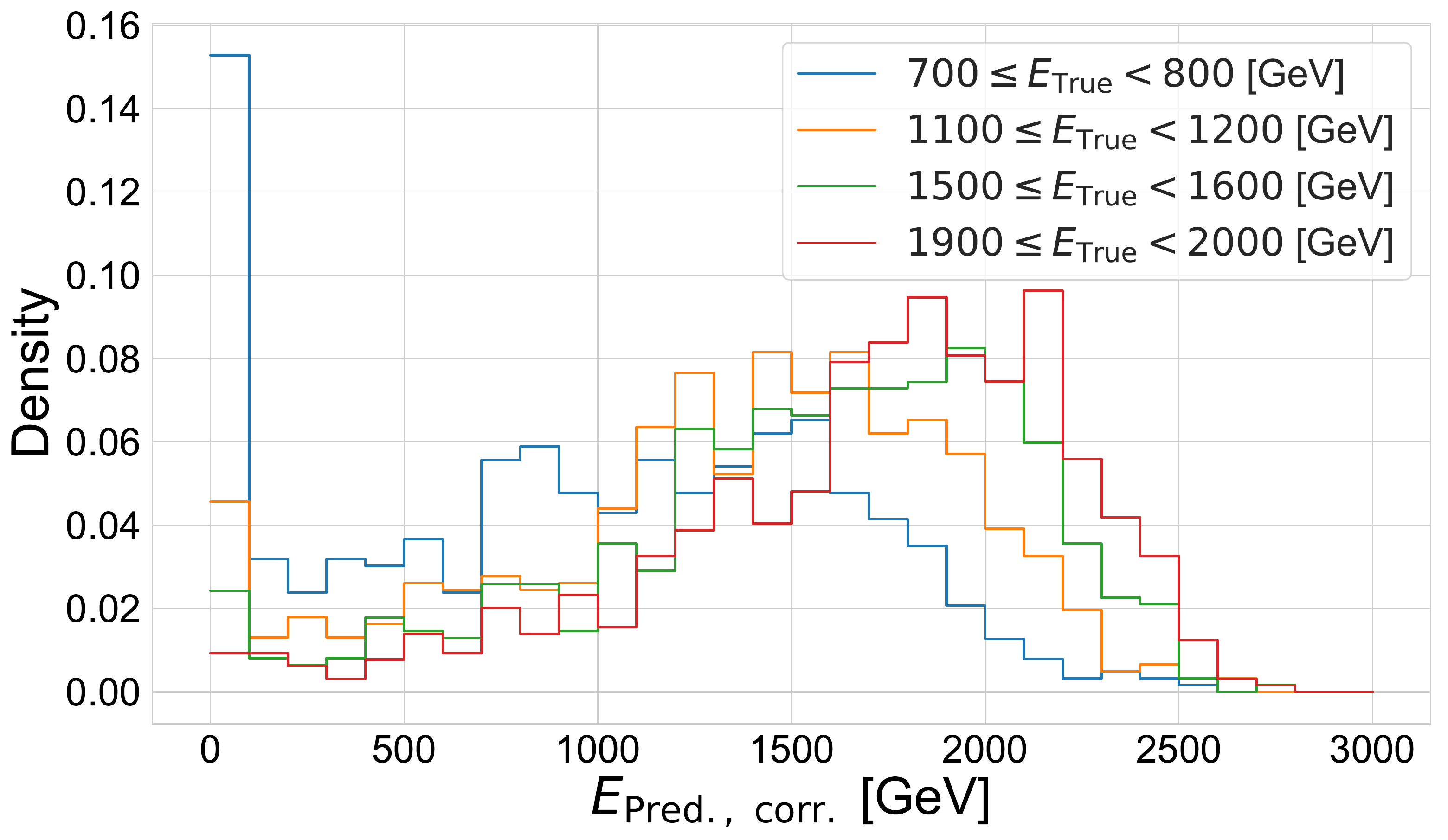}} %correctedmarginals.PNG}}
\caption{\em Distribution of predicted energy for the \kNN regressor in four bins of true energy.
}
\label{f:correctedmarginals}
\end{figure}
\end{center}

\begin{center}
\begin{figure}[h!]
\centerline{\includegraphics[width=0.9\textwidth]{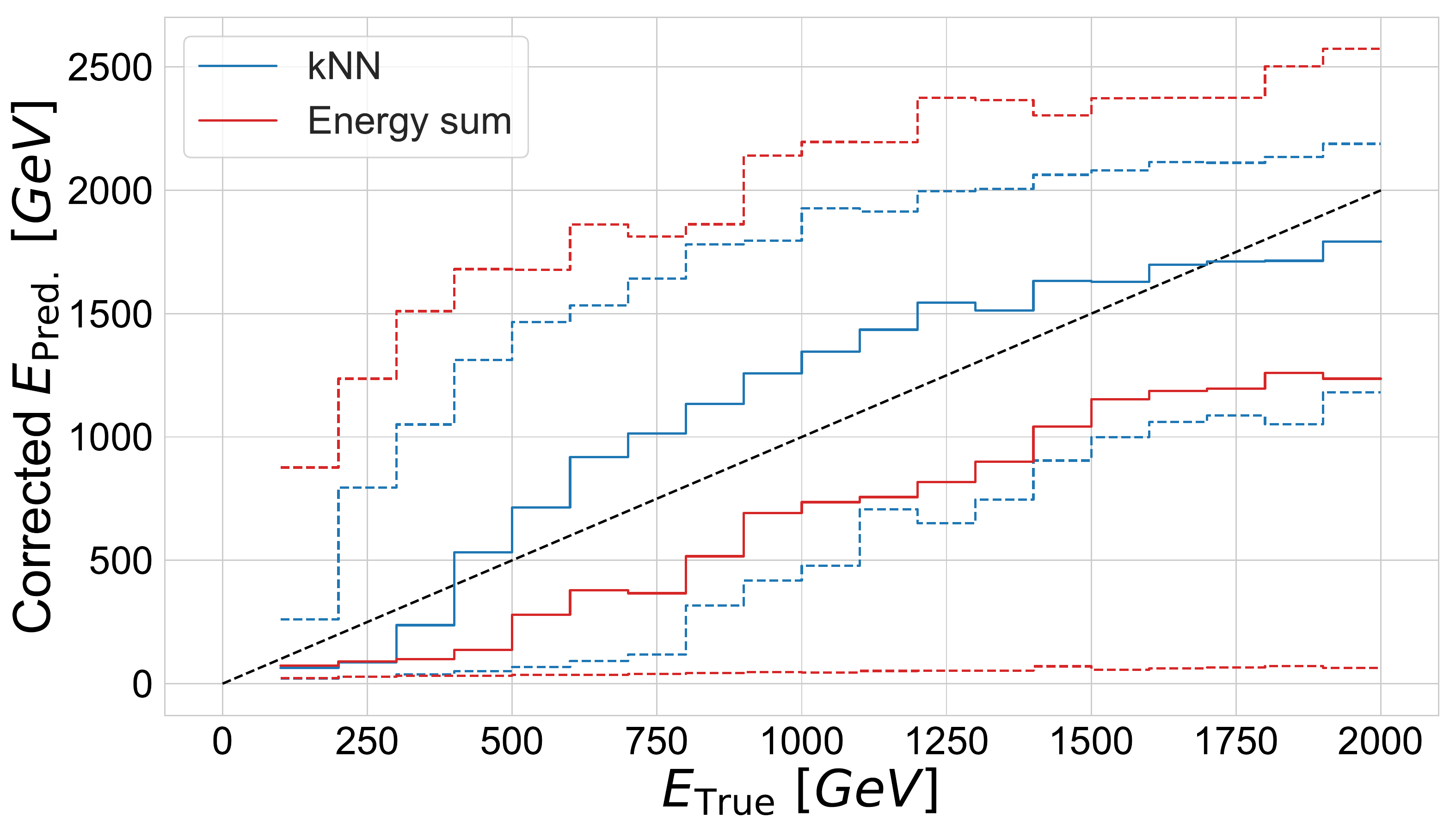}} %comparison.PNG}}
\caption{\em 68.3\% central intervals and 50\% percentile of the distribution of predicted energy as a function of true muon energy, for the \kNN regressor (blue) and for the energy sum model (red). A diagonal line (black) is also shown. The large improvement of the \kNN over the energy sum model is evident, as is its higher linearity. }
\label{f:comparison}
\end{figure}
\end{center}

\begin{center}
\begin{figure}[h!]
\centerline{\includegraphics[width=0.9\textwidth]{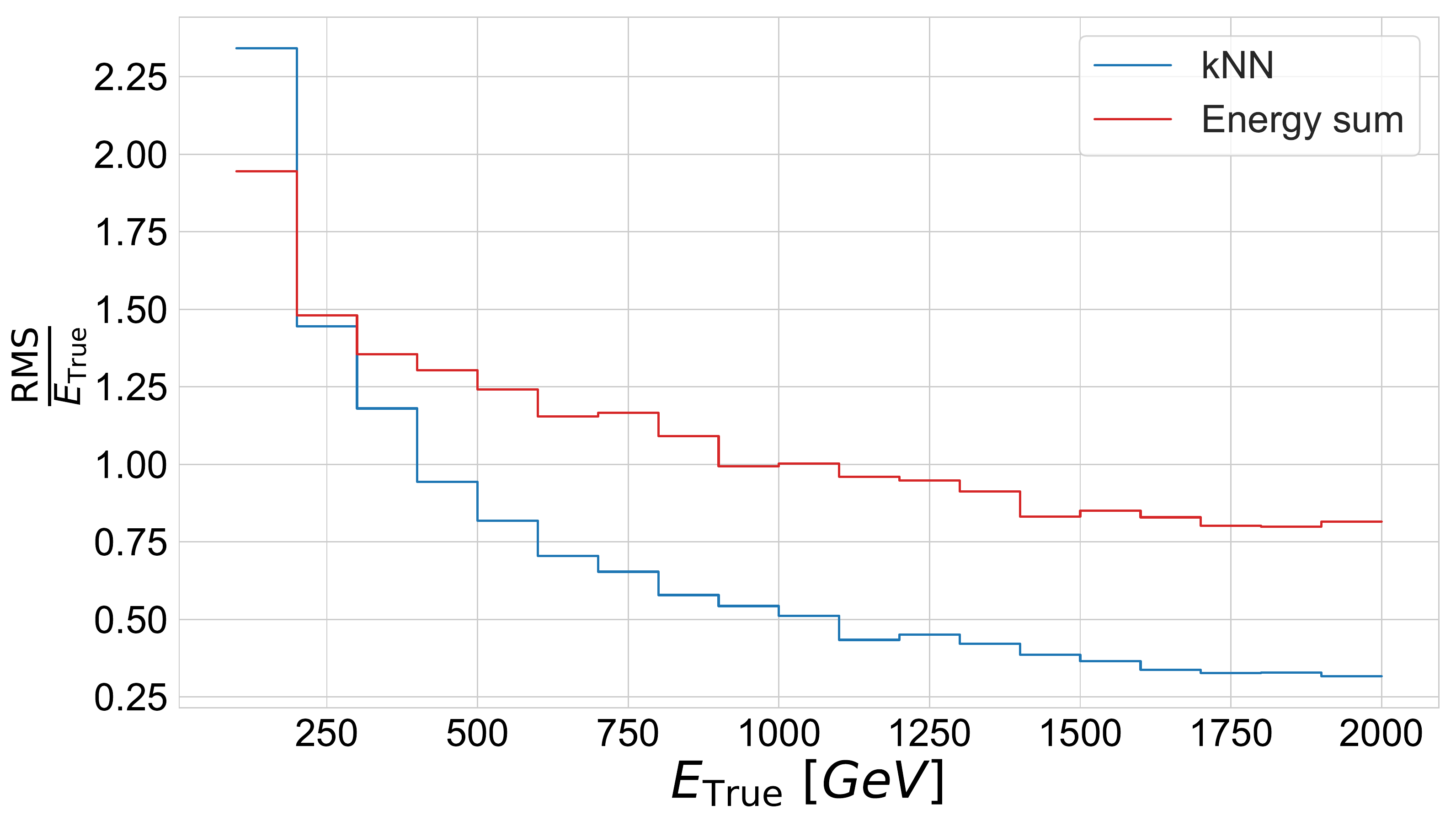}} %relrms.PNG}}
\caption{\em Relative RMS ($\sigma(E_{pred})/E_{pred}$) of the predicted muon energy as a function of true muon energy for the \kNN (blue) and the energy sum (red) models. }
\label{f:relrms}
\end{figure}
\end{center}

In Figures~\ref{f:uncorrectedkNN}, \ref{f:correctedkNN}, and  \ref{f:correctedmarginals} are presented the results of the \kNN regression, employing optimized weights and parameters mentioned above. A summary of the 68.3\% central intervals of predicted energy as a function of true energy for \kNN and energy-sum NN model is presented in \autoref{f:comparison}, showing that the spatial information summarized in the additional 15 features discussed in \autoref{s:features} may indeed be fruitfully exploited. Indeed, the combined \kNN regressor outperforms the energy sum model by a large amount. \autoref{f:relrms} shows the relative uncertainty in predicted muon energy as a function of true muon energy for the two models, further proving the point. 

\clearpage 
\section{Conclusions}\label{s:conclusion}

The particularities
of muon interactions with matter have until recently forced us to rely on magnetic bending for the measurement of the energy of muons. As we move towards
the investigation of the potential of machines envisioned by the recently published ``2020 Update of the European Strategy for Particle Physics"\cite{eusupp}, we must ask ourselves how we plan to determine the energy of multi-\si{\tev} muons in a particle detector operating at a higher energy collider than LHC. As mentioned {\em supra} (\autoref{s:introduction}), the CMS detector is able to achieve relative resolutions in the range of 6\% to 17\% at 1 TeV, thanks to its very strong 4-Tesla solenoid. This, however, relies on a compact design that may be suboptimal in future machines offering higher luminosity and collision energy. Given the linear scaling with momentum of relative momentum resolution determined with curvature fits, it is likely that magnetic bending alone is not the best answer. 

In this work we have investigated, using an idealized calorimeter layout, how the spatial information from energy deposits left by radiation emitted by very energetic muons may be exploited to obtain estimates of muon energy. In order to exploit the information contained in a set of 16 high-level features extracted from the energy and spatial information of the detected deposits, we have found useful to employ a $k$-nearest neighbor technique. Given the results of this initial investigation, it will be interesting to study whether further increases in performance may be obtained by deep learning methods employing, {\em e.g.,} convolutional neural networks.

The produced study shows that the fine-grained information on the location of the deposits allows one to significantly improve the precision of muon-energy estimates. For muons in the $[1.5,1,6]$ TeV range, for example, the use of all information results in a 68.3\% central interval estimate of $[904,2063]$ GeV (with 1633 GeV as 50th percentile); the use of only the total energy in a simple regression returns a central interval of $[70,2303]$ GeV (with 1042 GeV as 50th percentile) (see Fig.\ref{f:comparison}).

Our results also indicate that indeed, measurements that start to become competitive with those of magnetic bending (using as a benchmark the relative resolutions of LHC experiments, linearly extrapolated to higher energy as demanded by curvature uncertainty scaling) may be achieved for muons at the top end of the studied energy range, 1.5 to 2 TeV, which is of practical interest for future collider applications. We expect that above that value the strong increase in radiative losses makes this measurement technique even more proficuous.

\FloatBarrier

\end{document}